\documentclass[fleqn,usenatbib]{mnras}

\usepackage{newtxtext,newtxmath}

\usepackage[T1]{fontenc}
\usepackage{ae,aecompl}
\usepackage{mathtools}

\usepackage{graphicx}	
\usepackage{amsmath}	
\usepackage{amssymb}	
\usepackage{siunitx}
\usepackage{mwe}
\usepackage{subfig}
\usepackage[usenames,dvipsnames]{xcolor}

\newcommand{\ds}{$\, {\rm deg}^2$}
\newcommand{\angstrom}{\text{\normalfont\AA}}

\title[The rest-frame UV LF at $3.5<z<4.5$]{The rest-frame UV luminosity function at $z \simeq 4$: a significant contribution of AGN to the bright-end of the galaxy population}

\author[N. J. Adams et al.]{
N. J. Adams$^{1}$\thanks{E-mail: nathan.adams@physics.ox.ac.uk},
R. A. A. Bowler$^{1}$,
M. J. Jarvis$^{1,2}$,
B. H\"au\ss ler$^{3}$,
R. J. McLure$^{4}$, 
\newauthor A. Bunker$^{1}$,
J. S. Dunlop$^{4}$,
A. Verma$^{1}$
\\
$^{1}$Sub-department of Astrophysics, University of Oxford, Denys Wilkinson Building, Keble Road, Oxford OX1 2DL, UK\\
$^{2}$Department of Physics, University of the Western Cape, Bellville 7535, South Africa\\
$^{3}$European Southern Observatory, Alonso de Cordova 3107, Vitacura, Santiago, Chile\\
$^{4}$Institute  for  Astronomy,  University  of  Edinburgh,  Royal Observatory, Edinburgh, EH9 3HJ, UK\\}

\pubyear{2019}

\begin{document}
\label{firstpage}
\pagerange{\pageref{firstpage}--\pageref{lastpage}}
\maketitle

\begin{abstract}
We measure the rest-frame UV luminosity function (LF) at $z \sim 4$ self-consistently over a wide range in absolute magnitude ($-27 \lesssim M_{\rm UV} \lesssim -20$). The LF is measured with 46,904 sources selected using a photometric redshift approach over $ \sim 6$~\ds~of the combined COSMOS and XMM-LSS fields. We simultaneously fit for both AGN and galaxy LFs using a combination of Schechter or Double Power Law (DPL) functions alongside a single power law for the faint-end slope of the AGN LF. We find a lack of evolution in the shape of the bright-end of the LBG component when compared to other studies at $z \simeq 5$ and evolutionary recipes for the UV LF. Regardless of whether the LBG LF is fit with a Schechter function or DPL, AGN are found to dominate at $M_{\rm UV} < -23.5$. We measure a steep faint-end slope of the AGN LF with $\alpha_{AGN} = -2.09^{+0.35}_{-0.38}$ ($-1.66^{+0.29}_{-0.58}$) when fit alongside a Schechter function (DPL) for the galaxies. Our results suggest that if AGN are morphologically selected it results in a bias to lower number densities. Only by considering the full galaxy population over the transition region from AGN to LBG domination can an accurate measurement of the total LF be attained.

\end{abstract}




\section{Introduction}

The origins and subsequent evolution of galaxies at high-redshift remains a source of many unanswered questions and disputes in astrophysics. The rest-frame Ultraviolet (UV) Luminosity Function (LF) of galaxies (number density of galaxies as a function of intrinsic UV luminosity, defined here at $\lambda_{rest}=1500$\AA) is a fundamental observable that can be used to trace the time evolution of the overall galaxy population and the physical processes contained therein. The UV emission from these galaxies is a tracer of both young stellar populations from recent star formation and hot accretion disks around supermassive black holes \citep[e.g.][]{Wang1996,Koratkar1999,Adelberger2000}.
By comparing the expected LFs derived from cosmological simulations to observations, it is possible  to obtain insight into the physical processes occurring within galaxies, such as baryonic gas cooling, star-formation rates and feedback processes  \citep[e.g.][]{Dekel1986,Nagamine2001,Benson2003,Powell2011,Silk2012}. 
A significant source of feedback impacting the bright-end of the LF has been attributed to Active Galactic Nuclei (AGN), which are postulated to hinder the growth of the most massive and luminous galaxies 
via kinetic feedback from jets as well as radiative feedback \citep[e.g.][]{Cole2002,Benson2003,Begelman2003,Bower2006}. This results in a coupling of black-hole activity to the properties of the host galaxy \citep[e.g. the $M_{\rm BH}-\sigma$ relation; ][]{Ferrarese2000,Graham2011} and a predicted sharper high-luminosity turn-down of the LF \citep[e.g.][]{Ciotti1997,Silk1998,Kauffmann1999,Binney2004,Schawinski2007}. This introduces a quantitative measure of when in cosmic time AGN are governing the shape of the LF or the galaxy mass function and how their influence may vary over time. Observationally, bright galaxies ($L>L^*$; where $L^*$ is the characteristic luminosity of the knee in the LF) are rare however, and therefore constraining the bright end of the LF in the very high-redshift universe, and hence determining the impact of AGN feedback at this epoch, presents a challenge.

Both galaxies and AGN at high redshift can be selected via their strong Lyman-continuum break, a technique first implemented nearly three decades ago by \citet{Guhathakurta1990} and~\citet{Steidel1992} \citep[see also][] {Steidel1996}.
The rest-frame UV emission is shifted into the optical regime of the observer at $z \geq 2.5$ and the break in the continuum can be observed in broadband photometry. 
The break allows for an approximate redshift determination either from a selection procedure using colour-colour cuts or through the implementation of photometric redshift fitting, resulting in the selection of so-called `Lyman-break galaxies' (LBGs).
To-date, many thousands of high-redshift LBGs have been found within deep~\emph{Hubble Space Telescope (HST)} data~\citep[e.g.][]{Bouwens2015, Finkelstein2015}.
In addition, using shallower but much wider ground-based surveys it has been possible to place the first constraints on the very bright-end of the galaxy LF, and the transition into the AGN regime~\citep[e.g.][]{Bowler2014,Ono2017,Stevans2018}.

Despite these advances, debate has persisted over the time evolution and functional form with which to describe the rest-frame UV LF at $z > 3$. 
Traditionally, the LBG LF has been fitted with a three-parameter Schechter function \citep{Schechter1976}, which is described by a normalisation ($\Phi^*$), characteristic luminosity or absolute magnitude ($L^*$ or $M^*$) and a faint-end slope ($\alpha$).
With the advent of wider-area data however, several works have found evidence for a departure from this form at the bright-end, with a double-power law (DPL) providing a better fit to the data at $z \simeq 4$--$5$~\citep{Ono2017, Stevans2018} and $z \simeq 6$--$7$~\citep{Bowler2014,Bowler2015}.
The four parameter DPL fit allows for greater control of the slope at the bright-end ($\beta$), rather than the fixed exponential decline of a Schechter function.  
In parallel to discussions on the shape of the high-redshift LF is further debate on the redshift evolution of the parameters.
Some previous works using predominantly~\emph{HST} imaging have favoured a fixed characteristic absolute magnitude of $M^* \simeq -21$ \citep[e.g.][]{Bouwens2015,Finkelstein2015}, whereas other studies have suggested a rapid change in this parameter at $z > 5$ \citep[e.g.][]{Mclure2009,McLure2013,Bunker2010,Wilkins2011,Oesch2012,Schenker2013,Schmidt2014,Bowler2015,Stevans2018,Viironen2018}. 

In parallel to studies of the LBG LF, there have been continued efforts to pin down the AGN/quasar LF out to high redshift. While previous LBG and AGN studies have been undertaken separately, the advent of wide-area LBG studies and deeper, smaller-area AGN studies, has resulted in a convergence of the samples. Around $M_{\rm UV} \simeq -23.5$ the number density of the sources of the two populations at $z\simeq4$ are found to be comparable \citep{Stevans2018,Ono2017}. As the two populations can be selected in similar ways, contamination makes the derivation of the faint-end AGN slope and bright-end LBG slope difficult. Solving this problem is necessary to pin down the contributions of AGN and LBGs to the ionising background in the epoch of reionisation (EoR). In the process of measuring the AGN LF, debate has arisen between a number of studies over the steepness of the faint-end slope \citep[e.g.][]{Glikman2011,Masters2012,Giallongo2015,Ricci2017, Parsa2018,Akiyama2018, Boutsia2018, Stevans2018}. With power-law values for the faint-end slope ranging from $\alpha \simeq -1.3$ to $\alpha \simeq -2.0$, this can have a significant impact on whether AGN are capable of sustaining reionisation on their own at this epoch and on where within the LF the dominant contributor changes between AGN and LBGs. To fully understand the shape of the faint-end of the AGN LF and the bright-end of the LBG LF, it is now apparent that both populations must be considered together.

In this work we determine the rest-frame UV LF at $z \simeq 4$ using the deep optical/near-infrared dataset available in the ground-based Cosmological Evolution Survey (COSMOS) and~\emph{XMM-Newton} Large-Scale Structure (XMM-LSS) fields.
Combined, these fields provide $\sim 6$\ds~of data that enables the LF of $3.5 < z < 4.5$ galaxies and AGN to be measured precisely over a wide range in absolute UV magnitude ($-26 \lesssim M_{\rm UV} \lesssim -20$) using a coherent selection methodology.
In comparison to previous studies at this epoch~\citep[e.g.][]{Ono2017,Stevans2018}, the inclusion of photometric data through to the $K_s$-band enables us to determine photometric redshifts with increased accuracy (4 to 5 per cent outlier rate), and provides a more complete selection than that provided by colour-cuts (up to 90 per cent).

The paper is structured as follows. In Section~\ref{sec:data} we provide a description of the data used in this study. The resultant photometric redshift estimations from template fitting, sample selection and completeness simulations are described in Section~\ref{sec:photozs}. In Section~\ref{sec:LFs} we detail the final sample and method used for the measurement and fitting of the LFs and present the binned LF results. We then discuss the resulting functional fits
in Section~\ref{sec:discussion} in the context of past and present studies. We present our conclusions in Section~\ref{sec:conclusions}.
We assume a standard cosmology with $H_0=70$\,km\,s$^{-1}$\,Mpc$^{-1}$, $\Omega_{\rm M}=0.3$ and $\Omega_{\Lambda} = 0.7$. All magnitudes listed follow the AB magnitude system \citep{Oke1974,Oke1983}.

\section{Data}\label{sec:data}

In this study we use a multi-wavelength data set consisting of 13-band photometry, with optical wavelength data from the Canada-France-Hawaii-Telescope Legacy Survey (CFHTLS) and the HyperSuprimeCam Strategic Survey Programme \citep[HSC;][]{Aihara2014,Aihara2017}. Near-infrared data is provided by from the VISTA Deep Extragalactic Observations (VIDEO) survey \citep{Jarvis2013} and UltraVISTA \citep{McCracken2012} in the XMM-LSS and COSMOS fields respectively. The details of the available bands and depths for each field are provided in Table 1. The details of data/image post-processing are described in \citet{bowler2019lack}. Fields were selected on the basis of containing both wide ($\approx 6$ deg$^2$) and deep ($m_i > 25.4$ for $5\sigma$ detections) optical and near-infrared coverage. Successfully identifying $z \simeq 4$ galaxies requires deep imaging in the rest-frame UV band (shortward of the $i$-band in the observed frame) and in the surrounding bands in the optical and near-infrared to cover the Lyman and Balmer breaks of the galaxy spectrum, which are key features for determining photometric redshifts. We create object masks to exclude regions of missing or poor-quality data.
Predominantly this involved masking the halos of foreground stars, image artefacts and the field edges. This leads to a removal of approximately 5 per cent of the total area available.

\subsection{COSMOS}
\label{sec:maths} 

Covering $\sim 1.5$\,deg$^2$ to a $5\sigma$ depth of $m_i = 26.6$, the COSMOS field contains the deepest data in our study. The optical imaging comes from HSC observations in the $GRIZy$ filters over the entire field. Additional optical data is provided by the CFHTLS-D2 field, which covers the central $1$\,deg$^2$ of the field in the $u^*,g,r,i$ bands.
We use the UltraVISTA data release 3 (DR3) imaging in the near infrared ($Y,J,H,K_s$) over the full area used.

\subsection{XMM-LSS}

The XMM-LSS field is covered fully by optical imaging from HSC. One of the four pointings from HSC, which is centered on the Ultra Deep Survey from UK Infrared Telescope \citep[UKIDSS;][]{Lawrence2007}, is deeper than the rest of the field. In addition to this, CFHTLS-D1 provides deep optical imaging in 1\ds~of the field. As a result of this, the data can be grouped into three primary sub-fields of uniform optical data: `HSC-DEEP' with $m_i = 25.4$ of area $1.48$\,deg$^2$, `HSC-UDEEP' with $m_i = 26.3$ of area $1.92$\,deg$^2$ and `CFHTLS-D1' with $m_i = 26.43$ of area $0.97$\,deg$^2$. All magnitudes are listed as $5\sigma$ detection limits. We extract $u^*$ fluxes from the CFHTLS wide field survey in order to enable the whole XMM-LSS field to have full coverage in this filter. Near-infrared data is obtained from VIDEO and is uniform across the entire field. The total area of imaging that we utilise after masking is 5.88\,deg$^2$ over the two fields.

\begin{table}
    \centering
      \caption{Summary of the $5\sigma$ detection depths within the COSMOS and XMM-LSS fields. Depths are calculated in 2\arcsec ~diameter circular apertures, placed on empty regions of the image. Sources in the catalogues have a point-source aperture correction applied.
      The XMM-LSS field is split into three regions of 1.5\ds~corresponding to the VISTA VIDEO tiles, and ordered from low to high RA. 
      The first tile `XMM1' contains the deeper HSC pointing and the third tile `XMM3' contains the 1\ds~CFHTLS D1 field.}
    \label{tab:FiveSig}
    \begin{tabular}{cccccc}
    \hline
    Filter & COSMOS & XMM1 & XMM2 & XMM3 & Origin \\
    \hline
    $u^*$ & $ 27.0$   & $25.8$  & $25.8$   &$ 26.9$& CFHT\\
    $g$ & $ 27.1$   & --  & --  &$ 27.0$& CFHT\\
    $r$ & $ 26.7$  & --  & --  &$ 26.6$& CFHT\\
    $i$ & $ 26.4$  & --  & --  &$ 26.4$& CFHT\\
    $G$ & $ 27.2$  &$ 27.0$ &$ 26.4$ &$ 26.5$& HSC\\
    $R$ & $ 26.8$  &$ 26.5$ &$ 26.1$ &$ 26.1$& HSC\\
    $I$ & $ 26.6$  &$ 26.4$ &$ 25.4$ &$ 25.6$& HSC\\
    $Z$ & $ 25.9$  &$ 26.3$ &$ 24.6$ &$ 24.8$& HSC\\
    $y$ & $ 25.5$  &$ 25.6$ &$ 24.1$ &$ 24.1$& HSC\\
    $Y$ & $25.5$  &$ 25.2$ &$ 25.1$ &$ 25.2$& VISTA\\
    $J$ & $25.3$  &$ 24.7$ &$ 24.7$ &$ 24.7$& VISTA\\
    $H$ & $25.0$  &$ 24.2$ &$ 24.3$ &$ 24.3$& VISTA\\
    $K_{\rm s}$ & $24.8$  &$ 23.8$ &$ 23.9$ &$ 23.9$& VISTA\\
    \hline
    \end{tabular}
\end{table}

\section{Methods}\label{sec:photozs}
\subsection{Photometric redshifts}

In this section we outline the process used for estimating redshifts for the galaxies in our fields. With the wealth of multi-wavelength data available it is logical to make maximal use of the information available for each candidate galaxy. We thus elect to use a photometric redshift method to select our $z \simeq 4$ sample rather than traditional colour-colour cuts. Photometric catalogues were generated from our images using {\sc SExtractor} \citep{Bertin1996} in dual-image mode, with 2\,\arcsec \, diameter circular apertures for photometry. The selection band was the $i$-band, from which forced photometry was performed on the remaining bands. Catalogues were cut at the 5$\sigma$ detection limit of the $i$-band for the regions described in the previous section. All fluxes are corrected with a point-source aperture correction, where we model the point spread function (PSF) with {\sc PSFEx} \citep{Bertin2011} (See \citet{bowler2019lack} for details). Our observations are seeing limited ($\approx$ 0.7-0.8\,\arcsec full-width-half-maximum, FWHM) meaning the image PSFs are the dominant contribution to measured sizes at the redshift range of interest, with mean sizes of 2kpc or $0.3^{\prime\prime}$ \citep[e.g.][]{Huang2013}.

We estimate galaxy redshifts with the template fitting photometric redshift code {\sc LePhare} \citep{Arnouts1999,Ilbert2006}. This publicly available code operates by minimising the $\chi^2$ of galaxy spectral energy distribution (SED) templates fit to the multi-band photometry and uncertainties. We set the uncertainties of the photometry to a minimum of 5 per cent during the fitting process.

\begin{figure*}%
    \centering
    \hspace*{-1.9em} \subfloat[XMM-LSS]{{\includegraphics[width=1.12\columnwidth]{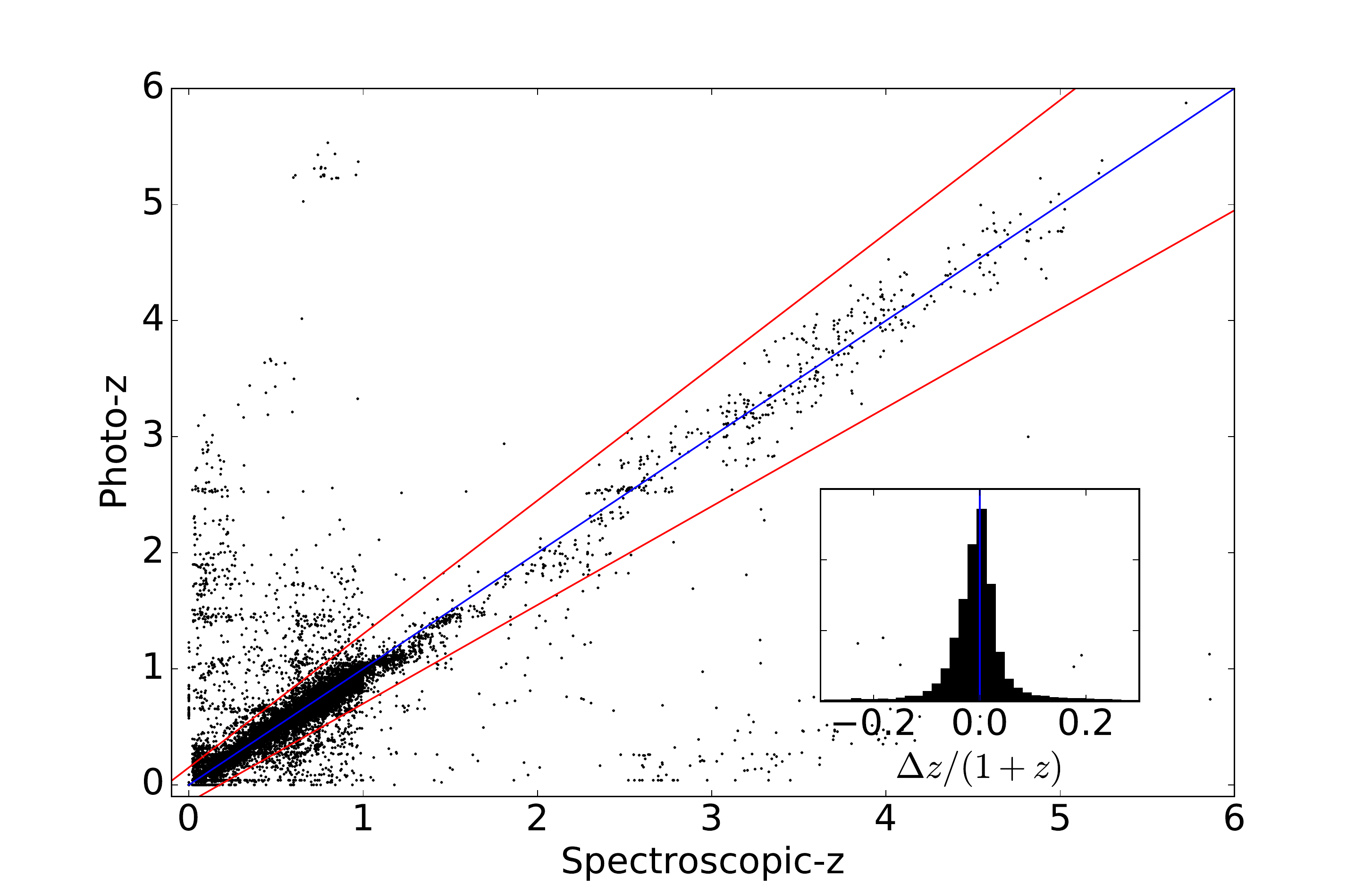} }} \hspace*{-4.9em}%
    \qquad
    \subfloat[COSMOS]{{\includegraphics[width=1.10\columnwidth]{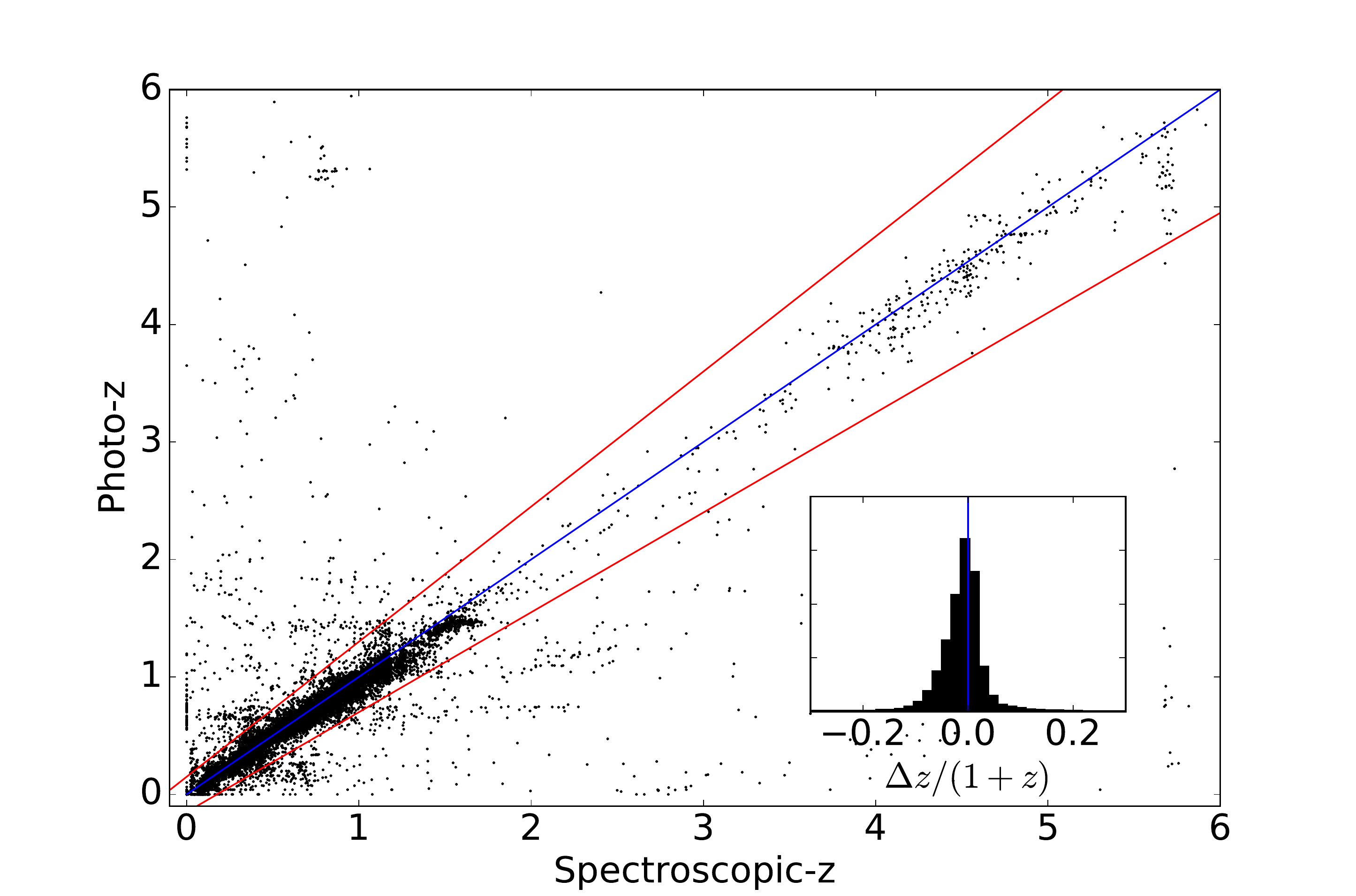} }}%
    \caption{A comparison of the photometric redshifts derived in this study to a compilation of spectroscopic redshifts. On the left we show the results for XMM-LSS and on the right, the results for COSMOS. The blue line shows the one-to-one correlation in the ideal case, and the upper and lower red lines show the 15 per cent margin in $1+z$ that defines a significant outlier. The sub-plot in each figure shows the histogram of the scaled photometric redshift deviations from the spectroscopic values. The XMM-LSS field includes an additional 455 high-redshift objects from VANDELS while COSMOS includes 3839 objects from DEIMOS. These were used to test photometric redshift outlier rates in the selection regime and not used in zero-point calibrations.}%
    \label{fig:PhotZ}%
\end{figure*}
 
\subsubsection{Template fitting}

{\sc LePhare} was run on all sources found by {\sc SExtractor} over the $\sim$6\,deg$^2$ of imaging. We used the COSMOS SED template set \citep{Ilbert2009}, where 32 templates are sourced from \cite{Polletta2007} with the GRASIL code \citep{Silva1998} and from \citet{Bruzual2003}. They cover a range of galaxy morphological classifications (E, S0, Sa, Sb, Sc, Sd,
Sdm) and have the necessary rest-frame wavelength range to cover our optical and near-infrared dataset. Within the fitting process, each of these templates is allowed to be modified for the effects of dust attenuation using the \citet{Calzetti2000} attenuation law and an attenuation value in the range $E(B-V) = 0-1.5$. At each redshift, we use the \citet{Madau1995} treatment for absorption by the inter-galactic medium (IGM). We experiment with the use of the model from \citet{Inoue2014} for the IGM in our photometric redshift methodology. The impact on the derived redshifts of our sample is a mean shift of 0.04 to higher redshift.  As the systematic shift is much less than the photometric redshift errors we derive ($\delta z \simeq 0.15$ at $z \simeq 4$), the results of this work are unchanged with the use of this alternative IGM model. Alongside this, template spectra for AGN from \citet{Salvato2009} and stars from \citet{Hamuy1992,Hamuy1994,Bohlin1995,Pickles1998,Chabrier2000} were fit and the $\chi^2$ statistics were used to apply initial object classification and contamination control.

\subsubsection{Zero-points}

One systematic that requires controlling is the photometric zero-points of the data. Zero-point errors can arise as a result of small systematics when modelling filter transmission functions, through biases within the choice of SED templates and from the calibration of the images. {\sc LePhare} is capable of deriving a correction for this by fitting galaxies that have known spectroscopic redshifts. Small corrections are made to the zero-point of each filter to optimise the results on the sample of galaxies with known redshifts. For our data, a sample of 22,652 galaxies with spectroscopic redshifts from XMM-LSS and 21,990 galaxies in COSMOS from a mixture of the VVDS \citep{LeFevre2013}, VANDELS \citep{McLure2018,Pentericci2018}, Z-COSMOS \citep{Lilly2009}, SDSS-DR12 \citep{Alam2015}, 3D-HST \citep{Skelton2014,Momcheva2016}, Primus \citep{Coil2011,Cool2013}, DEIMOS-10K \citep{Hasinger2018} and the FMOS \citep{Silverman2015} surveys were used. Only spectroscopic redshifts with flags indicating high-quality were used (confidence of $\geq 95$ per cent). The inclusion of a large number of galaxies from different surveys in the zero-point calculations minimises the risk of the correction containing biases due to the individual surveys and their sample selections. Once the corrections were obtained, the zero-points were set and {\sc LePhare} was re-run on the entire sample. The corrections themselves are minimal compared to the uncertainty in the photometry ($<0.08$ mag).

\subsubsection{Photometric redshift accuracy}

Using the spectroscopic catalogues we are also able to assess the accuracy of our photometric redshifts which we show in Fig.\ref{fig:PhotZ}. Following \cite{Ilbert2009} and \cite{Jarvis2013} we evaluate the accuracy using the Normalised Median Absolute Deviation \citet[NMAD;][]{hoaglin2000understanding}. In the case of photometric redshifts, this is $1.48 \, \times\,$median[|$\Delta z$|/$(1+z)$] and is used because it is resistant to extreme outliers. We define outlier photometric redshifts as being greater than 15 per cent different in $1+z$ to known spectroscopic redshifts. For our sample we find our photometric redshifts have an outlier rate of 5.4 per cent and a NMAD of 0.031 in the XMM-LSS field. For COSMOS we find a 3.9 per cent outlier rate and a NMAD of 0.027. Compared to results from \citet{Jarvis2013} with 3.3 per cent outlier rate and NMAD of 0.025 in XMM-LSS, our spectroscopic sample is significantly larger and extends across the whole field rather than just the deeper CFHTLS-D1 region, making it more representative of the whole field.

\subsection{Sample selection}\label{sec:Acc}

For our final $z \simeq 4$ sample we use galaxies that have the highest peak in their redshift probability distribution function within the redshift range $3.5<z<4.5$. We applied an initial  cut of $\chi^2$ $<1000$ in order to remove suspect contaminants which are poorly fit to our galaxy and AGN SED templates (removing the worst $\sim0.01$ per cent of the population). The application of these selection criteria provided an initial sample of 74,699 galaxies and AGN at $3.5 < z < 4.5$.

\begin{figure}
\centering
\includegraphics[width=\columnwidth]{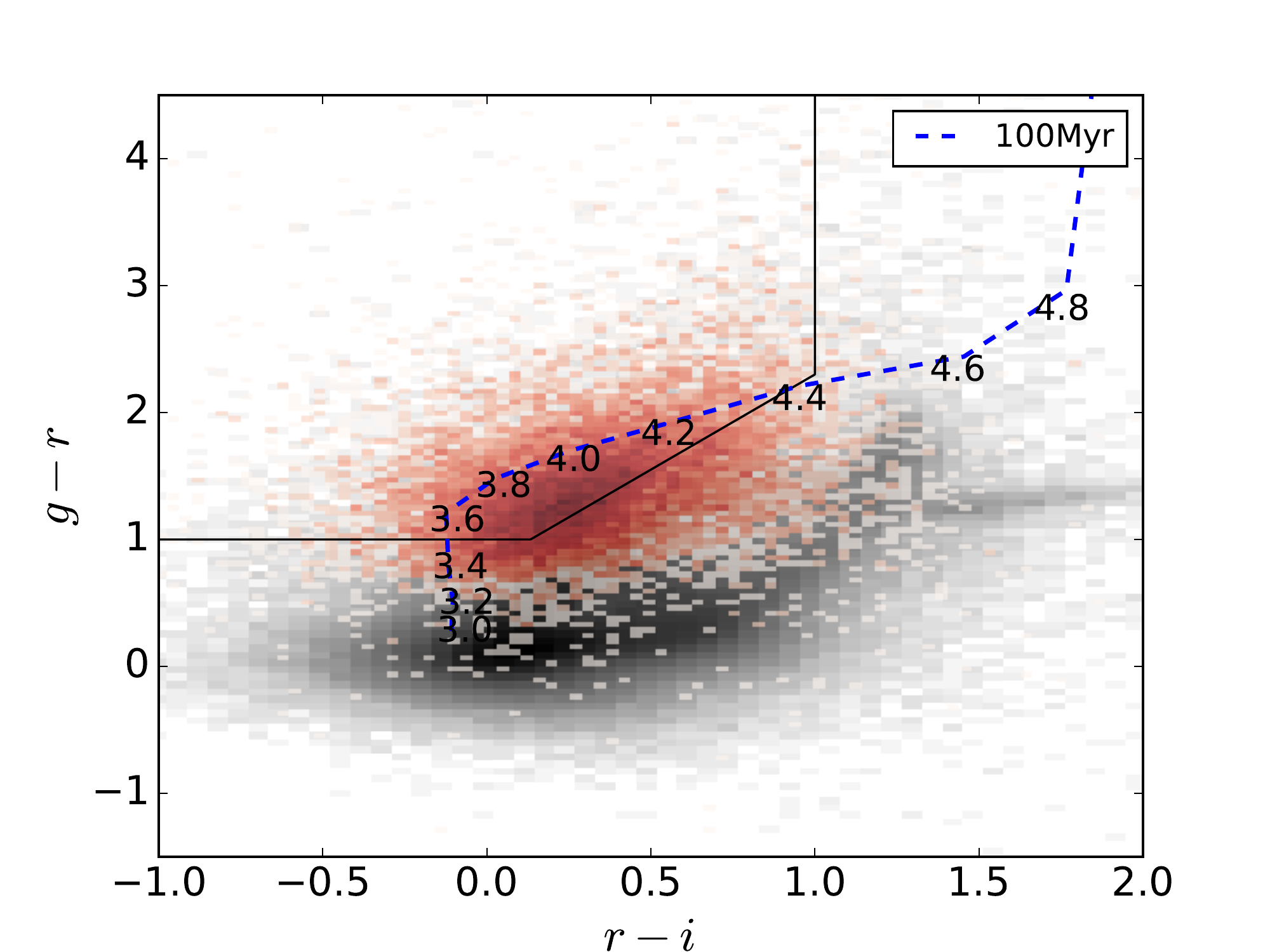}\caption{The $g$--$r$ against $r$--$i$ colour for our $3.5 < z < 4.5$ sample, selected using a photometric redshift analysis.
The background black distribution shows the complete catalogue used in our study, with the $3.5 < z < 4.5$ sample shown as the red distribution.
The colour-colour cut used by~\citet{vdb2010} is shown as the solid black line.
In comparison to our sample, this colour-cut is expected to be around 60 per cent complete at $z \simeq 4$.
An example galaxy evolutionary track from~\citet{Bruzual2003} is also displayed as the dashed blue line with the colour at each redshift labelled. The model has constant star formation, an age of 100\,Myr and $E(B-V)=0.15$.}
\label{fig:CCcut}
\end{figure}

Several previous studies have used colour-colour cuts to select LBG samples. While a conservative cut can provide a relatively pure sample of galaxies by taking advantage of the strong Lyman-break, they are often incomplete. In Fig.~\ref{fig:CCcut} we show the location of our $3.5<z<4.5$ sample in colour-colour space, along with the colour selection criteria used by \citet{vdb2010}. We find that 60 per cent of our candidate $3.5<z< 4.5$ galaxies would have been identified with such a colour-colour selection, in line with the 55--65 per cent completeness estimated by \citet{vdb2010}. Also shown is an example evolution track of a \citet{Bruzual2003} model galaxy, which passes through our sample over the desired redshift range as expected.

To examine our ability to successfully identify $z\simeq4$ objects we take a closer look at galaxies with a spectroscopic redshift in the redshift range of $3.5 < z < 4.5$ in the VANDELS and DEIMOS surveys. These were not used in the zero-point calibrations to avoid any favourable bias. We find a photometric redshift success rate in the sub-fields ranging from 87 to 90 per cent for the VANDELS (HSC-UDEEP) and DEIMOS (COSMOS) spectroscopic samples respectively. Half of the objects with an incorrect photometric redshift have bimodal solutions with the secondary solution being within the correct redshift region. Objects from the VANDELS survey are within our `HSC-UDEEP' sub-field and have faint infrared magnitudes ($24 < m_H < 26$). This is thus an indication of the worst case within our data as the deep $i$-band in this sub-field selects these faint objects while the relatively shallow $u^*$ and infrared coverage provides lower confidence of the detection of the Lyman break and galaxy continuum. We thus expect the completeness value of 87 per cent to be a lower bound of the photometric redshift completeness in our sample. Contamination from known low redshift objects ($z<2$) having photometric redshifts measured within $3.5<z<4.5$ is found to be very low ($\leq 0.05$ per cent). It is currently difficult to completely understand how many contaminant objects reach the final sample. With the simple assumption that the measured rates of contamination from the spectroscopic sample can be expanded to the full photometric sample, this would lead to a contamination rate of low redshift objects in the final $z\simeq4$ sample to be of order 1-2 per cent.

To assess the prospects of classifying AGN candidates and to further test the robustness of our photometric redshifts we match our COSMOS catalogue to the 16 objects identified spectroscopically as AGN at $z \sim 4$ in \citet{Boutsia2018}. As four of these objects are located out of the UltraVISTA field, we expect twelve of these objects to be located in our catalogues. As expected, all twelve are found in our catalogues. Ten are found to have the correct redshifts within errors and seven of those are found to have a better match to an AGN template than a galaxy template. We examine the SED fits for the two misclassified objects. We find that both have a suitable AGN fit at the correct redshift range, however a star template is found to have a lower $\chi^2$ when fit to our photometry. One of these objects is heavily blended with a bright star (we account for these effects in our area and completeness calculations in section 3.3), while the other is located in a small region with no $u^*$-band data where the photometric redshifts are consequently expected to be of a slightly lower quality.

\subsection{Selection completeness}\label{sec:completeness}

In order to assess the completeness of the source extraction processes we simulated our selection procedure by injecting 3.6 million fake sources into our images and recovering these using {\sc SExtractor}.
Our simulated galaxies have an assumed disk profile, with a fixed Sersic index of $n=1$ \citep{1963BAAA....6...41S,Conselice2014}.
For fainter magnitudes ($M_{\rm UV} > -22.5 $) the UV luminosity of each fake galaxy is drawn from a posterior distribution that assumes the LF Schechter parameters and redshift evolution from \citet{Bouwens2015}. This ensures we appropriately account for Eddington bias. The distribution is flat for brighter galaxies ($M_{\rm UV} < -22.5 $) where the effect of this scattering is much weaker due to the higher confidence detections. 
The half-light radii of the galaxies are selected from the galaxy Size-Luminosity distribution of \citet{Huang2013}. From this distribution the mean half-light radii of LBGs at $z\sim 4$ are generally smaller than 2kpc or $0.3^{\prime\prime}$, therefore the measured sizes in our imaging are dominated by the seeing. 
Simulated galaxies were sampled onto the pixel scale of the images, convolved with our model for the point-spread function and scaled to the corresponding flux for the randomly assigned absolute magnitude.

\begin{figure}
\centering
\includegraphics[width=\columnwidth]{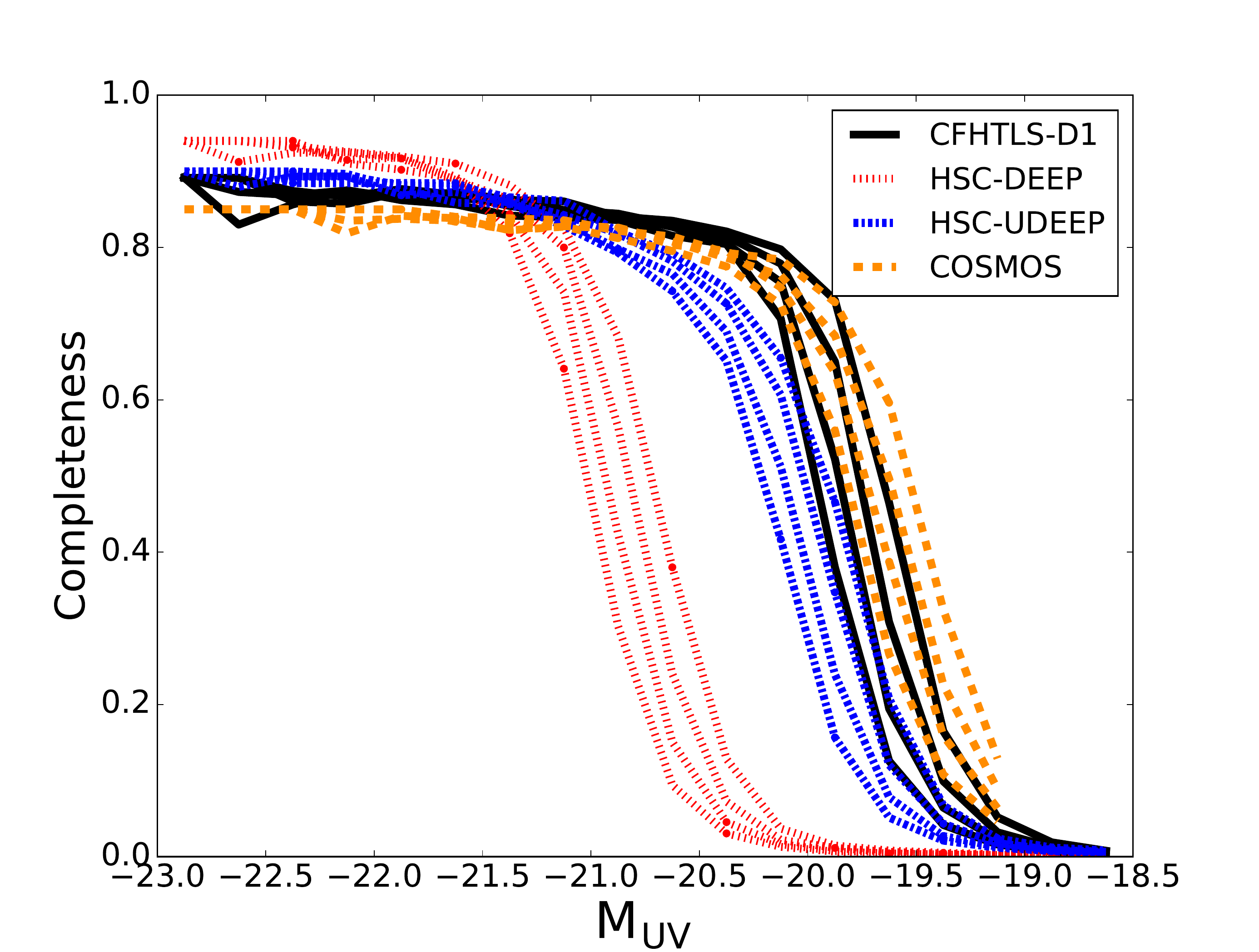}\caption{Completeness curves for our selection criteria as a function of absolute UV magnitude for the four subfields. Each subfield is represented by the colours black for CFHT-D1, red for HSC-DEEP, blue for HSC-UDEEP and yellow for COSMOS. For each sub-field we show four lines representing the variability of completeness across the full redshift bin $3.5 < z < 4.5$. Each line is a bin of 0.25 in redshift and from left to right is high redshift to low redshift, higher redshift bins being less complete due to the impact of distance on apparent luminosity. Deeper fields also appear more crowded, so have a higher probability of blending and hence converge to lower values of completeness at the bright-end.}
\label{fig:Comp}
\end{figure}

To avoid unrealistic blending or image crowding, we added 2000 fake sources for each run in the CFHT images, and 4000 for the larger HSC images.
To account for blending of sources, we assume that all fully-blended sources are unrecoverable. Our simulated galaxies are prevented from being inserted into our images where the segmentation maps from {\sc SExtractor} show the locations of pre-existing objects. This restriction applies to the central location of the simulated galaxy, so instances of partial blending ($<50$ per cent overlap) are still present and accounted for. Due to the prevention of total blending in the simulations, our completeness curves converge to a value of 1.0 at the bright-end. These lines are then scaled down by the fraction of the field which is occupied by pre-existing sources in order to account for the probability a source is lost to total blending. This is equivalent to reducing the total surveyed area by the proportion occupied by foreground objects. The corrected completeness functions then converge to the ratio of empty field area to total field area. The deeper the field the more dense it appears and so deeper fields converge to lower completeness values (see Fig.~\ref{fig:Comp}). 

Simulated galaxies were placed into the image as faint as $M_{\rm UV} = -18.5$ in order to account for objects scattering above and below our selection limits in apparent magnitude.
Each sub-field has its completeness function measured in bins of 0.25 in redshift in order to take into account changes in apparent luminosity with distance. Each redshift bin has a total of 200,000 simulated sources inserted into the images, totalling 3.6 million simulated objects across all sub-fields. 
We measured the completeness as the ratio of retrieved galaxies from {\sc SExtractor} to the total number of galaxies injected into the simulation as a function of absolute magnitude.
In Fig.~\ref{fig:Comp} we show the completeness curves for the different regions of imaging we defined.
The results of these simulations were used to set the faintest magnitude used in this work.
We only present the LF in luminosity bins where the completeness is $> 50$ per cent in at least two sub-regions across the redshift range, to ensure that the contribution from each sub-field to the LF is not dominated by the completeness correction and to minimise cosmic variance (described in more detail in Section 4.1).

\section{The Luminosity Function}\label{sec:LFs}

The result of our photometric redshift selection procedure is a sample of 74,699 LBGs and AGN at $3.5 < z < 4.5$ from the XMM-LSS and COSMOS fields. Using this sample we proceed to measure the rest-frame UV LF at $z \simeq 4$.

\subsection{The 1/Vmax method}

We use the 1/$V_{\rm max}$ \citep{Schmidt1968,rowanrobinson1968} method to measure the LF. The maximum observable redshift of each object was found by taking the best fit SED and redshift from {\sc LePhare} and subsequently redshifting this further in steps of $\Delta z = 0.01$ to find the maximum redshift at which it would drop below the flux limit of our sample in each field. This is performed while accounting for the different $i$-band depths of each sub-field. The value of $V_{max}$ then corresponds to the co-moving volume between the maximal observable redshift of the objects and a lower bound of $z=3.5$. The final rest-frame UV LF, $\Phi(M_{\rm UV})$, for our complete sample of galaxies is then determined using:
\begin{equation}\label{eqn:lf}
\Phi(M) d \log(M) = \frac{1}{\Delta M } \sum_i^N \frac{1}{C_{i,f} V_{max,i}} ,
\end{equation}
where $\Delta M$ is the width of the magnitude bins and $C_{i,f}$ is the completeness correction for a galaxy $i$ depending on its location within the sub-fields, $f$. 

We estimate the uncertainty of the LF in each magnitude bin using:

\begin{equation}
\delta \Phi(M) = \frac{1}{\Delta M} \sqrt{\sum_i^N \left(\frac{1}{V_{max,i}}\right)^2} .
\end{equation}

In the lower luminosity bins we adopt a bin size in absolute magnitude of $\Delta M = 0.25$. As we move towards brighter magnitudes, where we begin to probe the AGN LF beyond $M_{\rm UV} = -24$, we increase the bin size to first $\Delta M = 0.5$ and then a maximum of $\Delta M = 1.0$ if the number counts drop below 10. This is performed in order to maintain good number statistics in each bin. We determine the UV absolute magnitude $M_{\rm UV}$ using a top-hat filter of width \SI{100}{\angstrom}, centred on \SI{1500}{\angstrom} in the rest frame, using the best-fitting SED model for each object. After removing all objects in bins with less than 50 per cent completeness, we compute the final LF with a total of 46,904 galaxies and AGN.

\subsubsection{Cosmic variance}

\begin{figure*}
    \centering

    \includegraphics[width=0.9\textwidth]{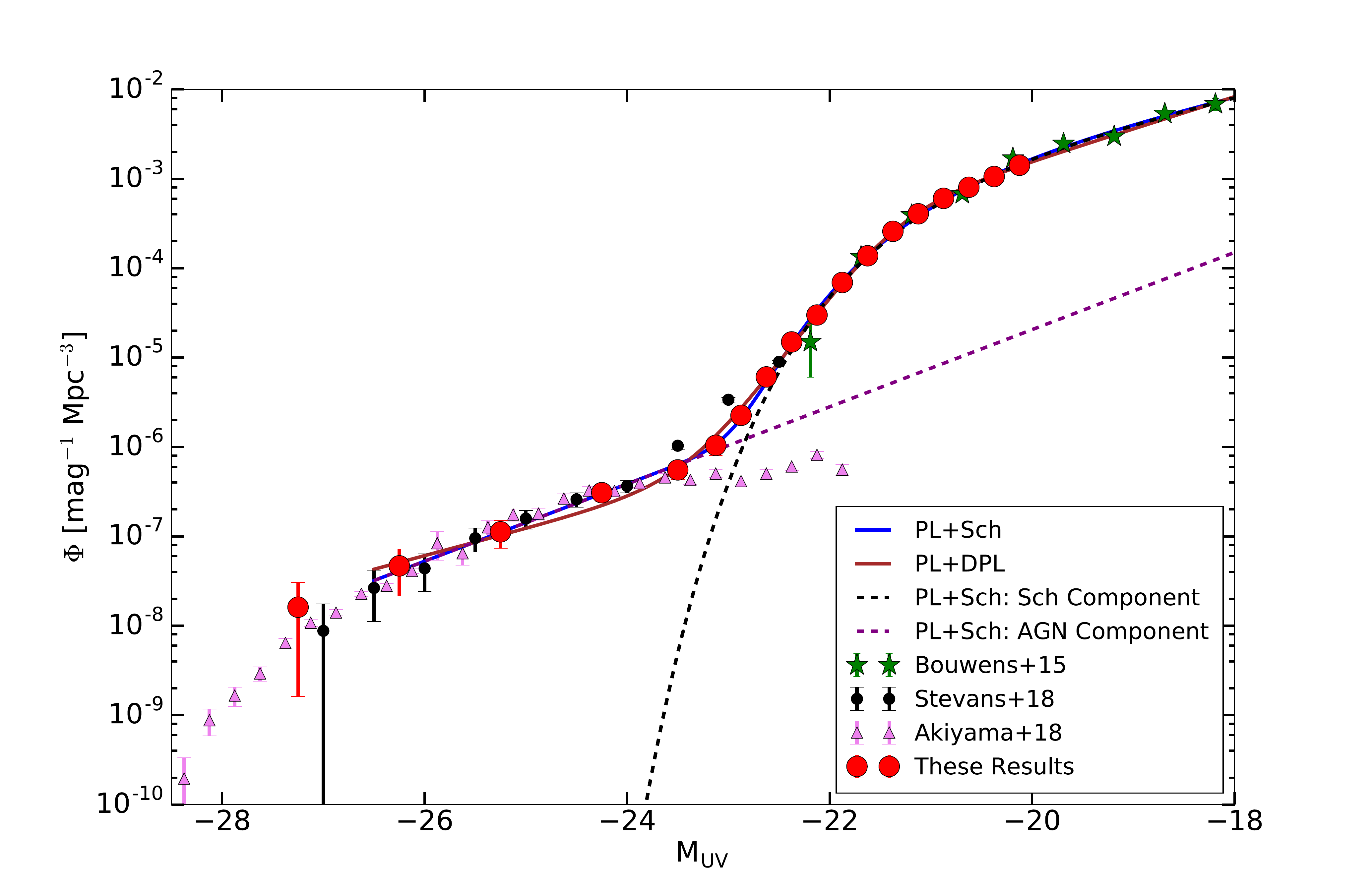}
    \caption{The $3.5<z<4.5$ UV LF derived in this study (red circles). The blue and red solid lines show the simultaneous fit of the AGN and LBG LF with a Schechter and DPL functional form for the LBGs respectively. The dashed lines show the individual AGN (purple) and LBG (black) contributions to the total LF for the PL+Sch case, combining to form the blue solid line. Shown alongside are results from previous studies by \citet{Bouwens2015} (green stars), \citet{Stevans2018} (black circles) and \citet{Akiyama2018} (violet triangles).}
    \label{fig:FULL}
\end{figure*}

\begin{table*}
\label{tab:FinalResult}
\centering
\caption{The results of fitting to the $3.5<z<4.5$ rest-frame UV LF derived in this study.
The first column lists the fitting parameterisation, with the next two columns showing the best-fit parameters for the AGN power law.
Columns 4--8 show the LBG fitting results, either for a Schechter or DPL.
The final columns shows the $\chi^2$ and reduced $\chi^2$ of the fit.
The top two rows show the results of simultaneous fitting of the AGN and LBG components, where we fit a single power law for the AGN component and either a Schechter function or DPL for the LBGs.
The bottom two rows show the results of fitting only the LBG component, where we include only data points fainter than $M_{\rm UV} \geq -23$. 
In all cases, at fainter than $M_{\rm UV} \geq -20$ we include the~\citet{Bouwens2015} points in the fitting which reduces errors by a factor of $\sim2$ by alleviating degeneracies on $\alpha$. 
The AGN normalisation is calculated at $M_{\rm UV}=-25.7$ and the AGN LF does not make use of the data point at $M_{\rm UV}=-27.25$.}
\begin{tabular}{ccccccccc}
\hline
                    & \multicolumn{2}{c}{AGN}                             &          \multicolumn{4}{c}{LBGs}                               &      \\
          & $\textrm{log}_{10}(\Phi)$ & $\alpha$  & $\textrm{log}_{10}(\Phi)$ & $M^*$ & $\alpha$ & $\beta$ & $\chi^2$ & $\chi_{\rm red}^2$ \\ 
           & $[\textrm{mag}^{-1} \textrm{Mpc}^{-3}]$ & & $[\textrm{mag}^{-1} \textrm{Mpc}^{-3}]$ & $[\textrm{mag}]$ & & & & \\  \hline

PL\textsubscript{AGN}+Sch    & $-7.15^{+0.21}_{-0.35}$                            & $-2.09^{+0.32}_{-0.38} $                    &  $-2.79^{+0.08}_{-0.08} $       & $-20.89^{+0.12}_{-0.10}$      & $-1.66^{+0.13}_{-0.08}$                      & --     & 25.8 & 1.36 \\
\\[-1em]
PL\textsubscript{AGN}+DPL   & $-7.07^{+0.29}_{-0.22}$                            & $-1.66^{+0.29}_{-0.58} $                    &  $-3.30^{+0.08}_{-0.06} $       & $-21.37^{+0.08}_{-0.11}$      & $-1.92^{+0.07}_{-0.04}$                      & $-4.92^{+0.29}_{-0.25}$     & 25.3 & 1.41 \\
\\[-1em]
Sch           & --                    & --                   & $-2.85^{+0.06}_{-0.06} $         & $-20.98^{+0.05}_{-0.06}$       & $-1.71^{+0.06}_{-0.05}$                      & --       &     37.2 &  2.33      \\
\\[-1em]
DPL          & --                    & --                   &   $-3.27^{+0.10}_{-0.08}$        & $-21.34^{+0.10}_{-0.12}$      & $-1.89^{+0.07}_{-0.06}$                       & $-4.77^{+0.22}_{-0.20}$   &   19.9  & 1.27 \\

\hline
\end{tabular}
\end{table*}

\begin{figure*}
    \centering
    \includegraphics[width=1.0\textwidth]{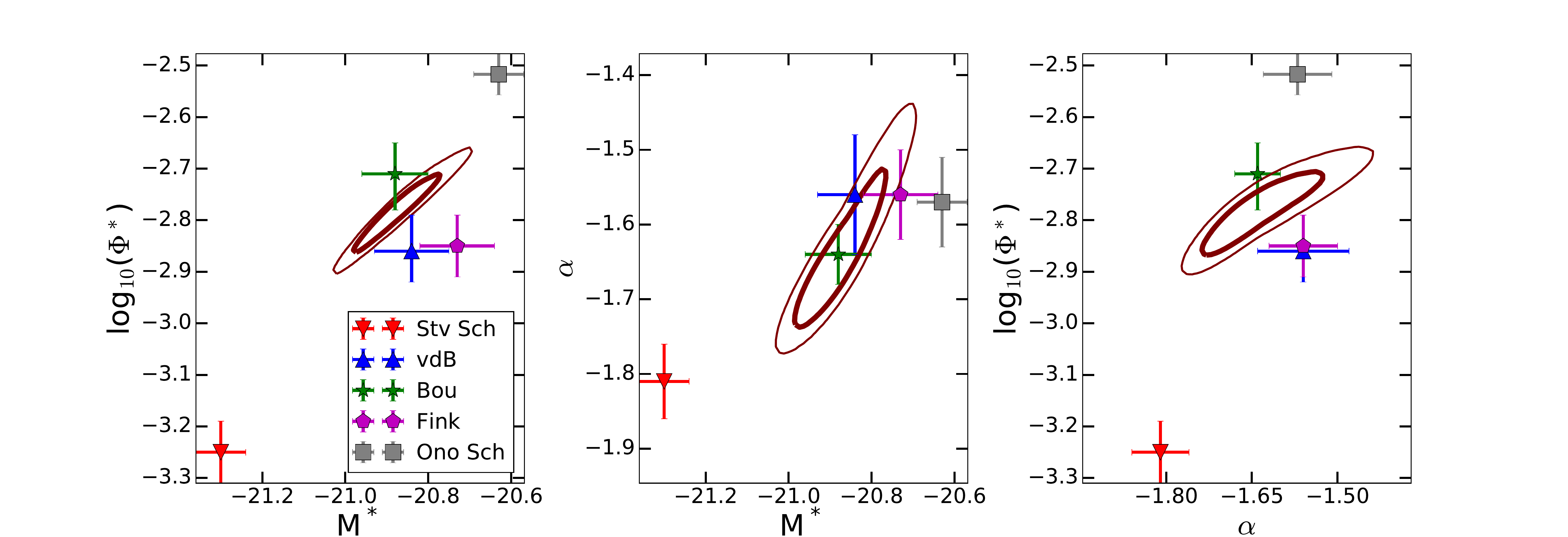}
    \caption{Contour plots showing the 1$\sigma$ and 2$\sigma$ confidence intervals for the LBG LF derived from our PL+Sch fit presented in Table 2. We compare our results to past studies that used the same functional form \citep[displayed as vdB, Fink, Bou, Ono and Stv respectively]{vdb2010,Finkelstein2015,Bouwens2015,Ono2017,Stevans2018}}
    \label{fig:Contour1}
\end{figure*}

\begin{figure*}
    \centering
    \includegraphics[width=1.0\textwidth]{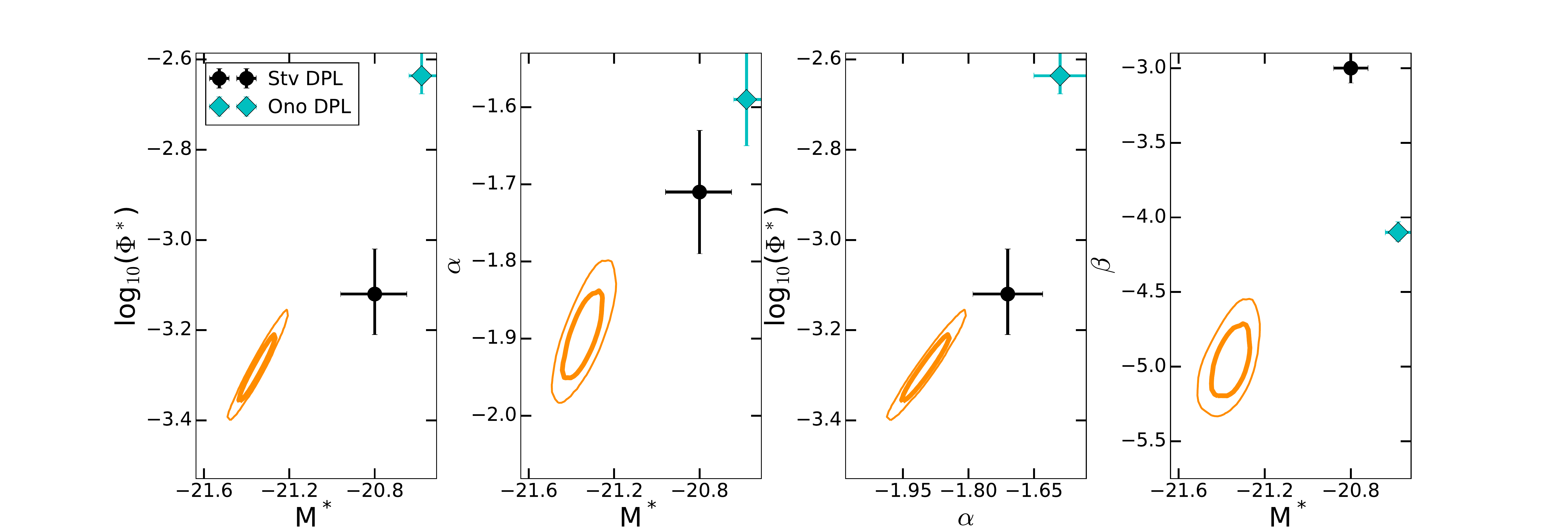}
    \caption{The same as Fig.\ref{fig:Contour1}, but for the PL+DPL fit. We compare to the results of past studies using the same functional form for the LBG LF. \citep[displayed as Ono and Stv respectively]{Ono2017,Stevans2018}}
    \label{fig:Contour2}
\end{figure*}

With the LF fundamentally being a measurement of galaxy number densities, it is prone to the influence of large scale structure in the Universe when small survey volumes are used. The presence of voids, clusters and filamentary structures can skew LF results derived from single, small fields. Such an effect is commonly referred to as `cosmic variance'. Making use of the \citet{Trenti2008} cosmic variance calculator with our galaxy samples, field areas and LF binning we find that the error due to cosmic variance is consistent at the 3.5-4.5 per cent level across the whole LF. This is caused by the competing effects of low number density at brighter luminosities and the reduction in survey volume at lower luminosities, as fields fail to be 50 per cent complete. To be conservative we adopt a 5 per cent cosmic variance error across all bins, added in quadrature to our LF uncertainty from Equation 2.

\subsubsection{Functional forms of the fits}

Using broad-band photometry in our study, it is not possible to robustly separate $z \sim 4$ LBGs from AGN dominated systems, and hence our sample bridges both populations (see Section \ref{sec:Acc} where 7 out of 10 spectroscopically confirmed AGN are successfully classified as such). Due to the depth of our data, we do not fully constrain the very faint end of the LBG LF at $M_{\rm UV} \geq -20$. To compensate we elect to use the data points from~\citet{Bouwens2015} for $M_{\rm UV} \geq -20$ to allow for stronger constraints on the LBG faint-end slope. 
Furthermore, although our sample contains objects as bright as $M_{\rm UV} \sim -26$, we do not have sufficient survey volume at this redshift to determine the position of the knee of the AGN LF at $M_{\rm UV} \simeq -26.5$ \citep{Akiyama2018,Stevans2018}. We therefore fit the AGN component of the LF with a single power law with the normalisation calculated at $M_{\rm UV} = -25.7$. 
We fit the data using different combinations of Schechter and DPL functions for the LBG component in addition to the single power law for the AGN component up to $M_{\rm UV} = -26.25$. When performing the fits, we make use of the Levenberg-Marquardt algorithm \citep{Levenberg1944,marquardt:1963} from {\tt scipy} to obtain a first order estimation of the fit parameters and their uncertainties. We then make use of a multi-dimensional $\chi^2$ grid to derive final uncertainties and contours for the fit parameters.

\subsection{The binned rest-frame UV LF}

The completeness-corrected rest-frame UV LF from our sample is shown in Fig.~\ref{fig:FULL}, with the full list of binned data points given in Appendix A.
Our results extend over a wide range in absolute UV magnitude, from the LBG dominated regime at $M_{\rm UV} = -20$, to the AGN dominated regime at $M_{\rm UV} \simeq -26$ and covers both the AGN-LBG transitional regime at $M_{\rm UV} \simeq -23.5$ and the LBG `knee' at $M_{\rm UV} \simeq -21$. 
We present our results alongside examples of past results covering the full luminosity range.
The studies of \citet{Akiyama2018} and \citet{Bouwens2015} cover the AGN and LBG luminosity functions respectively and the work by \citet{Stevans2018} covers a similar, but slightly brighter, UV luminosity range to this work. Our measurements of the LF are consistent with these past studies at both ends of our luminosity range. At $M_{\rm UV} \simeq -23.5$ we find a lower number density of objects when compared to those from \citet{Stevans2018}, a feature we explore in more detail in Section 5.2.

The best-fit parameters for the Schechter and DPL functional forms for the LBG LF, alongside the power-law slope for the faint-end AGN, are shown in Table 2. Our best performing functional form (minimal $\chi^2_{\rm red}$) to describe the LBG population is the Schechter function by a small margin, although a DPL is also acceptable. In Fig.\ref{fig:Contour1} and Fig.\ref{fig:Contour2} we present the $1\sigma$ and $2\sigma$ contours of the fit parameters for the Schechter and DPL parameters respectively for the LBG LF. These are both presented alongside the results from a small selection of past studies which have used the same functional forms to describe the $z\simeq4$ LF. More detailed discussion on the comparisons between the studies is presented in Section \ref{previous}.

In addition to the fits with the AGN LF included, we also show in Table 2 the results for fitting a Schechter and DPL function to just the data points fainter than $M_{\rm UV} = -23$, the magnitude where AGN contamination begins to become influential. This is performed to replicate the conditions of an LBG study that doesn't consider any significant AGN contamination in its fitting procedures and to observe what the consequences of this method would be. We find that the changes to the fits, when compared to our fits inclusive of AGN, are minimal and less than the $1\sigma$ level. However, there is a significant change to the $\chi^2$ values for the fits where the DPL functional form stands out as being significantly better. This is likely driven by AGN beginning to contribute to the number counts at $M_{\rm UV} \sim -23$.

\begin{figure*}
    \centering
    \includegraphics[width=0.90\textwidth]{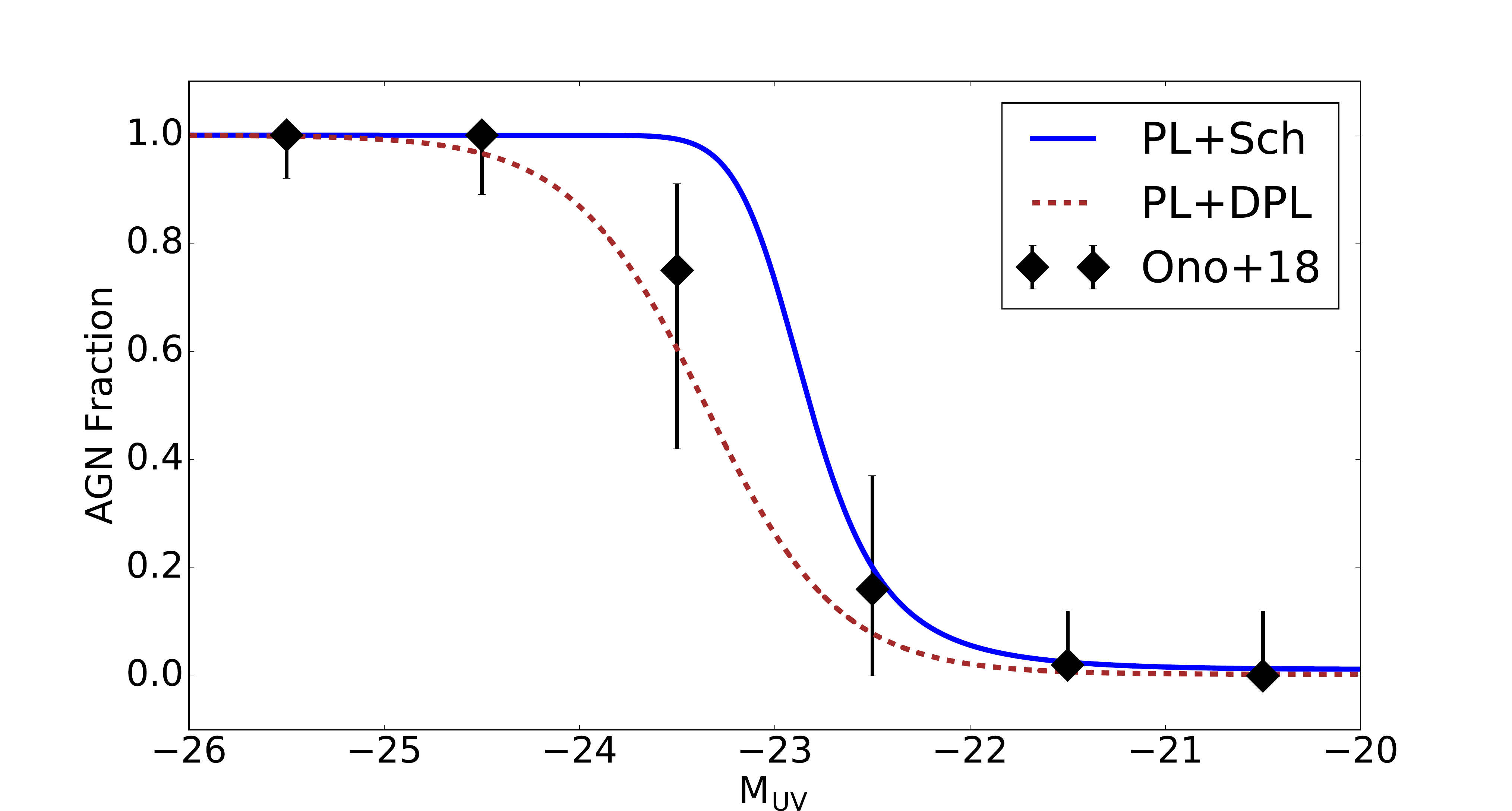}
    \caption{The fraction of objects estimated to be AGN as a function of absolute luminosity, found by taking the ratio of the best-fit AGN and LBG LF from Table 2. Also shown are the AGN fractions measured by \citet{Ono2017} who conducted a spectroscopic campaign of a subset of the objects contained within their sample.}
    \label{fig:AGNF}
\end{figure*}

As the LBG LF approaches $M_{\rm UV} \simeq -23$ from the faint-end, the fraction of objects which are AGN begins to rise~\citep[e.g.][]{Bian2013,Ono2017}. 
When attempting to determine the functional form of the LBG LF, these AGN can act as contaminants, resulting in an excess at the bright-end of the LBG samples.
If unaccounted for, this `AGN contamination' results in the measurement of a shallower slope of the bright-end of the LBG LF when using the DPL functional form.
In Fig.\ref{fig:AGNF} we show the fraction of objects that are AGN as a function of absolute UV magnitude derived from our best fit functions for both LBGs and AGN in Table 2, along with the results from \citet{Ono2017} that probed this regime with a spectroscopic sample. In both cases our results show that AGN make a significant contribution to the UV LF as faint as $M_{\rm UV} \simeq -23.0$. We note that the DPL fit has a smoother transition from LBG to AGN dominance than the Schechter function, which has an exponential cut off. At $M_{\rm UV} \simeq -23$ there is a very large disparity between the two functional forms, with an AGN fraction in the range $\sim$ 20--70 per cent depending on whether a DPL or Schechter function is used (see Fig.\ref{fig:AGNF}).

\section{Discussion}\label{sec:discussion}

\subsection{The LBG LF at $\mathbf{z=4}$}\label{previous} 

\begin{figure}
    \centering

    \includegraphics[width=\columnwidth]{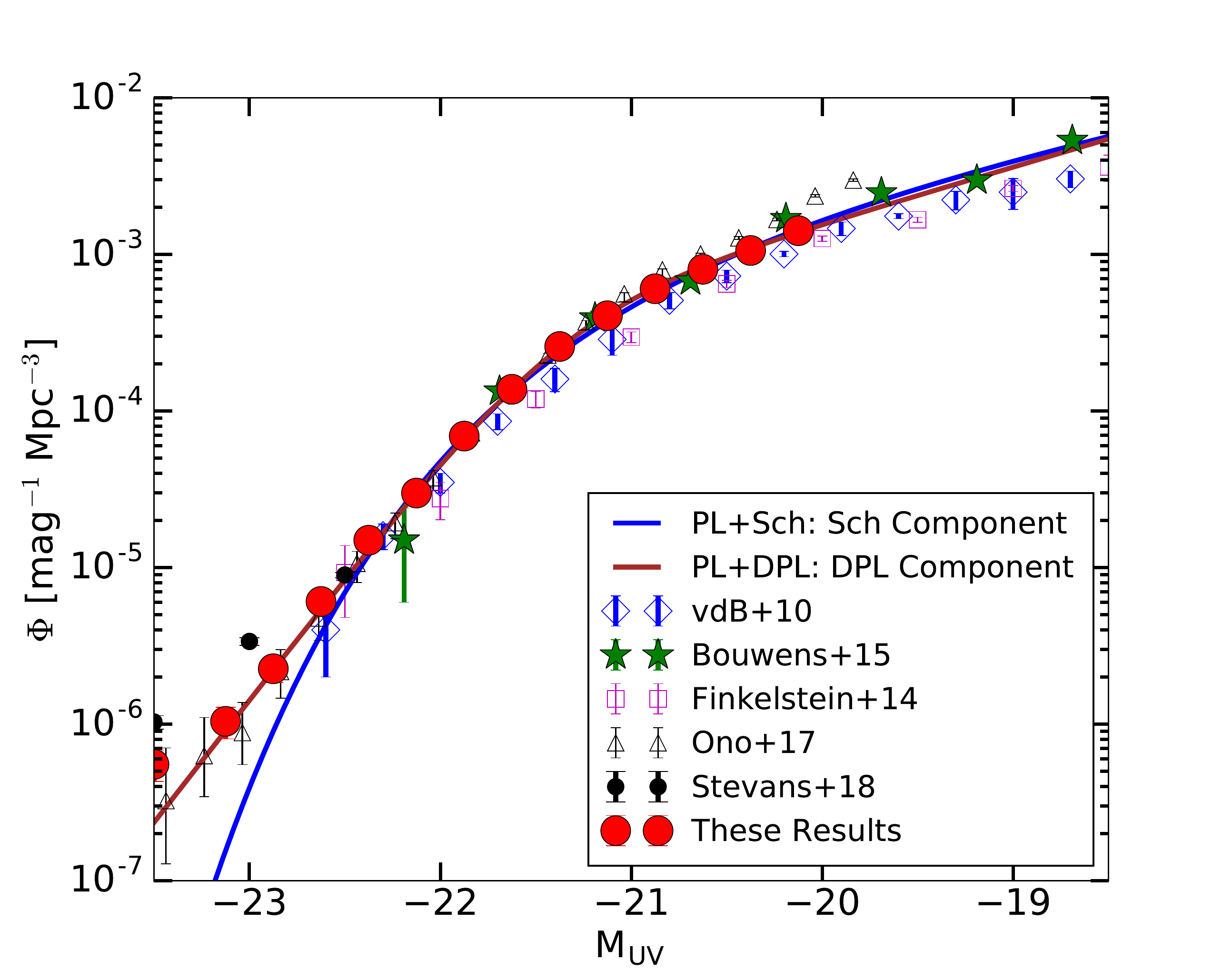}
    \caption{A zoom into the faint-end $M_{\rm UV} >= -23$ of the $z\sim4$ LF. The blue/brown lines show the LBG component Schechter/DPL fits that are fit alongside AGN. Shown alongside are a myriad of past studies targeting the $z\sim 4$ population \citep{vdb2010,Finkelstein2015,Bouwens2015,Ono2017,Stevans2018}.  At high luminosities, our results and the study by \citet{Stevans2018} are inclusive of AGN while \citet{Ono2017} attempted to remove them.}
    \label{fig:faintend}
\end{figure}

In Fig.~\ref{fig:faintend} we show our results faintward of $M_{\rm UV} = -23$ and compare to the results of previous studies of $z \simeq 4$ LBGs. We find good agreement with the results of~\citet{Bouwens2015} and~\citet{Ono2017}.
Fainter than $M_{\rm UV} \geq -22$ we find that the studies of~\citet{vdb2010} and~\citet{Finkelstein2015} show a lower number density than this work. \citet{Stevans2018} suggest that the~\citet{Finkelstein2015} sample may be lower than other studies due to the inclusion of~\emph{Spitzer} data in the mid-infrared, which assists in the removal of additional contaminants such as faint Milky Way brown dwarfs. We expect our photometric redshifts, which make use of full near-infrared coverage from VISTA and higher signal-to-noise ratio cuts, to also be robust to contamination from brown dwarfs. It is presently not clear why there is this offset between studies at the fainter-end of the LBG LF and more work is required to determine the cause.

Around $M_{\rm UV} \sim 22.5$, where the contribution of AGN begins to have an effect, we can compare our results to two previous studies.
\citet{Ono2017} used HSC imaging to select a sample of LBGs and corrected for AGN contamination through their spectroscopic measurements presented in Fig. \ref{fig:AGNF}. Their final LBG LF is found to agree well with our DPL fit at the bright end ($M_{\rm UV} < -21$).
We also find lower number counts in the $-24<M_{\rm UV}<-22.5$ regime than~\citet{Stevans2018}, who utilised 18\ds~of optical imaging from the Spitzer-HETDEX Exploratory Large-Area Survey \citep[SHELA;][]{Papovich2016}.
Within this magnitude range, our completeness is high, and the inclusion of two fields reduces the impact of cosmic variance.
The deep homogeneous multi-wavelength data gives us confidence in our results in this regime. We discuss the origin of this difference in Section \ref{sec:AGN}

In Fig. \ref{fig:Contour1} we compare the best-fit values and uncertainties for our Schechter fits against the aforementioned past studies. In this case our best-fit values are more closely matched to those of \citet{Bouwens2015} (within $ 1 \sigma$ for all parameters) and centralised in the spread of the other results. It is because of this close matching to \citet{Bouwens2015} that we elect to use their data points to constrain the far faint-end $M_{\rm UV} > -20$. By relieving degeneracies on $\alpha$, introducing these points reduces the errors on the best-fit LBG parameters by up to a factor of two while maintaining the same best-fit values to within $\sim1\sigma$. The result from \citet{Ono2017} provides a value of $\alpha$ which is in reasonable agreement with the studies of \citet{vdb2010},\citet{Finkelstein2015},\citet{Bouwens2015} and our own results, which are all in general agreement with each other. The degeneracies between $M^*$ and $\Phi^*$ may be responsible for the slight disagreement in $M^*$ and $\Phi^*$ measured by \citet{Ono2017}. The resultant best-fit Schechter function from \citet{Stevans2018} produces values for all three parameters which are clear outliers to the collection of other past results; we believe this to be driven by their excess in the total number density around $-24 < M_{\rm UV} < -23$ (See Section 5.3 and Fig.\ref{fig:FULL}).

For the DPL fits in Fig.\ref{fig:Contour2} there are fewer studies to compare against, however, we find we disagree with both past studies that have attempted a DPL fit \citep{Stevans2018,Ono2017}. Although the linear trend observed in the two-dimensional contour plots show the large impact of degeneracies the DPL parameters can have. Compared to \citet{Stevans2018} the differences are primarily driven by our much steeper bright-end slope which drives $M^*$ to brighter values. Our DPL model is found to fit very well to the AGN corrected data points at the bright-end of the LBG LF from \citet{Ono2017} (see Fig.~\ref{fig:faintend}). The differences in the best-fit parameters are in this case primarily driven by the excess of objects that are found by \citet{Ono2017} at $M_{\rm UV} > -21$ that leads to a fainter derived $M^*$. At brighter UV luminosities ($M_{\rm UV} < -24$) our measured AGN number densities are found to closely match those of \citet{Akiyama2018} and \citet{Stevans2018}.

\subsection{Evolution of the rest-frame UV LF}
 
 We compare our results to the two linear redshift evolution models presented in \citet{Bouwens2015}, derived from a mixture of results in the redshift range $4 < z < 8$. One model has a mildly-evolving value of $M^*$ and one has a fixed value of $M^*$, justified by the evolving model producing such a shallow slope that $M^*$ does not evolve with significance over the redshift range in which the model was created. We find agreement with the predictions to within $\sim 1\sigma$ for both the evolving and non-evolving $M^*$ models respectively, suggesting little to no evolution of the shape of the bright-end of the LBG LF. This is unsurprising given our close matching to their original data and their $z \sim 4$ bin being their most constrained and hence most influential on any evolutionary fit attempted.
 
 We also compare our results to simple evolutionary models presented by \citet{Bowler2015}. Within \citet{Bowler2015} a strong evolution in the Schechter function value of $M^*$ is found from $z=7$ to 5. Our measured value of $M^*$ is $-20.89^{+0.12}_{-0.10}$ which is 0.2 magnitudes fainter than their $z \sim 5$ measurement of $M^*=-21.07^{+0.09}_{-0.09}$ and much fainter than predicted if we extrapolate their proposed evolution to $z=4$ ($M^*\sim-21.3$), suggesting a non-linear/slowing evolution. When compared to the DPL results in \citet{Bowler2015} we find a strong agreement with the shape of the bright end of the $z \sim 5$ fit, with their measured $M^*=-21.40^{+0.13}_{-0.12}$ and $\beta=-4.8^{+0.3}_{-0.4}$ all agreeing within $\sim 1\sigma$. Regardless of whether we assume a Schechter or DPL functional form (both are plausible with our data), these results suggest little to no evolution in the bright end from $z\simeq 4-5$. 
 
\subsection{Discrepancies in the AGN/LBG transitional regime}\label{sec:AGN}

When considering the results of the DPL fits, the overall normalisation of the LBG fits have values similar to those measured by \citet{Stevans2018}. However, a greater discrepancy is observed at the high-luminosity end where we observe a much steeper bright-end slope and brighter turn off with $\beta = -4.92^{+0.29}_{-0.25}$ and $M^* = -21.37^{+0.08}_{-0.11}$ with our DPL fit. \citet{Stevans2018} and \citet{Ono2017} find values of $\beta = -3.8$, $M^* = -20.8$ and $\beta = -4.1$, $M^* = -20.58$ respectively for the two parameters. While the discrepancies with \citet{Ono2017} can be somewhat explained by their use of colour-colour selection and the uncertainty in AGN contamination, the very clear discrepancy with \citet{Stevans2018} (see Fig.\ref{fig:faintend}) can be explained by examining the differences in the methodology between our two studies. 

The selection criteria used in \citet{Stevans2018} is more complex than our own in an attempt to dig deeper into their shallower data while maintaining purity. They use a $3.5\sigma$ detection limit and cuts in both colour space and the redshift probability distribution. However, we believe that the observed discrepancy arises due to the fact that their $M_{\rm UV}$ values are calculated by converting the measured $i$-band apparent magnitude directly into an absolute magnitude using the estimated redshifts of each object. While the $i$-band contains the rest-frame 1500\AA, its positioning within the filter is redshift dependant and the actual flux relative to the average $i$-band measurement is dependant on the UV slope of the galaxy SED which can vary from source to source. The use of the $i$-band as a proxy for $M_{\rm UV}$ also leaves individual objects vulnerable to larger noise fluctuations in the $i$-band due to the lower confidence detections. These sources of potential inconsistency are accounted for in our methodology through the use of $5\sigma$ detection limits and the use of the best fit SED plus top-hat filter for calculating $M_{\rm UV}$. This makes our measurement of the absolute magnitude consistent between all galaxies and not as heavily reliant on the measurement in a single band, as the best-fit SED is constrained by all of the surrounding bands.

If we remeasure our luminosity function using the measured $i$-band flux as a proxy for $M_{\rm UV}$ we find a general trend of galaxies up-scattering to brighter bins. Galaxy numbers in every bin increase and the largest relative effect is found to be in the region where the LF is at its steepest (at $M_{\rm UV} \simeq -23$). When measuring $M_{\rm UV}$ in this way, the transition at $-24 < M_{\rm UV} < -23$ closely matches the results of \citet{Stevans2018} (see Fig.\ref{fig:MUV} in the Appendix). This highlights the importance of taking full consideration of the shape and redshift of the galaxy SED when measuring $M_{\rm UV}$.

\subsection{The $\mathbf{z \sim 4}$ AGN LF}

A subject of ongoing debate with regards to the AGN LF at high redshifts is the steepness of the faint-end slope. \citet{Akiyama2018} measures a relatively flat faint-end slope of $\alpha = -1.3$, while in this study we measure $\alpha = -2.09^{+0.32}_{-0.38}$/$-1.66^{+0.29}_{-0.58}$ respectively for the Sch/DPL fits. This is in agreement with \citet{Stevans2018}, however our AGN slope is slightly steeper when the DPL functional form is used for the LBGs. Between $-26 \leq M_{\rm UV} \leq -24$ the measured LF of the three studies agree very well, it is only when the transition to the LBG LF is reached that discrepancies arise. Both our work and the work of \citet{Stevans2018} make use of full template fitting and have data stretching into the near-infrared while the \citet{Akiyama2018} results are limited to just optical observations from HSC, resorting to stringent colour cuts and constraints on object morphology to minimise contamination. Through calculating $M_{\rm UV}$ with the $i$-band measurement as a proxy, the results from \citet{Stevans2018} show a greater number density of objects in the steep part of the LF in the transitional regime (see Section 5.3). Consequently, this leads to a shallower LBG bright-end slope, higher overall inferred galaxy fractions and a shallower AGN slope when using a DPL fit.

Another recent study by \citet{Boutsia2018} targeted the $z\sim 4$ AGN LF using an X-ray selected spectroscopic sample at $-24.5 < M_{\rm UV} < -23.5$. They find number densities of AGN to be high in this regime and similar to the values measured by \citet{Stevans2018}, leading to their conclusion that the total UV LF should be dominated by AGN at $M_{\rm UV} < -23.5$. In Fig.~\ref{fig:AGNF} we show that both functional forms of our measured LBG LF give a high AGN fraction within this range of absolute luminosities and may be the dominant population in the LBG selection to absolute magnitudes as faint as $M_{\rm UV}=-23$, in agreement with \citet{Boutsia2018}.

Purely photometric studies that focus on the AGN LF and rely on colour-colour selection and morphology (sources with a PSF-like profile) leave open the possibility of missing some AGN sources. Where the transition of the AGN to LBG luminosity function occurs, sources with weak AGN will have more significant contributions from the host galaxy. This can impact the measured profile of the source. Depending on the data used, selection criteria and extraction method, this could cause some objects to be missed through misclassification, leading to an underestimation of the AGN LF in such studies and making the completeness more challenging to model.

Within the results from the photometric study conducted by \citet{Akiyama2018}, which uses morphology in AGN classifying, a shallow faint-end slope of $\alpha = -1.3$ is measured. However, when a cut is made to their data brighter than $M_{\rm UV}= -23$, $-23.5$ and $-24$ mags, where we now know the transition from AGN to LBG LFs occurs, this value steepens to $\alpha = -1.57$, $-2.07$ and $-2.21$ respectively and more closely matches the results of this study and that of \citet{Stevans2018}. Other studies using additional selection criteria to extract pure AGN samples, such as X-ray emission \citep[e.g.][]{Giallongo2015,Parsa2018}, tend to have poorer number statistics and so uncertainties in the calculated faint-end slope of the AGN LF remains larger than the discrepancy between the various studies and provide no additional constraining power. Examining Chandra X-ray data available in COSMOS \citep{Marchesi2016} we recover the \citet{Boutsia2018} sample along with an extra source at $z_{phot} = 4.36$ which is outside of the selection range of $3.6<z<4.2$ implemented by that study. At $M_{\rm UV} \lessapprox -23.5$, 66 per cent of our sources have an X-ray counterpart with that fraction dropping off rapidly at fainter UV magnitudes. Together, this highlights that measuring the faint-end of the AGN LF remains challenging at luminosities fainter than the transition into LBG dominance ($M_{\rm UV} \gtrsim -23$) and that the key to solving issues around the faint-end AGN slope relies on secure object classification with well understood completeness (examples including spectroscopic surveys, use of multiwavelength signatures and Baldwin, Phillips \& Terlevich diagrams; \citealp{Baldwin1981}).

\section{Conclusions}\label{sec:conclusions}

We exploit deep optical/NIR data from the COSMOS and XMM-LSS fields to measure the rest-frame UV LF at $z \simeq 4$. The combination of depth and area allows us to measure the LF from $-27 < M_{\rm UV} < -20$ using 46,904 objects selected through a photometric redshift method. Our conclusions on the resultant LFs are:

\begin{enumerate}
    \item When fit alongside AGN, we find we are unable to confidently discern between the two LBG functional forms of a Schechter function or DPL. When we fit an LBG LF to only those data points fainter than $M_{\rm UV} = -23$ we find that the DPL stands out as being the better descriptor. However, this is mostly driven by the inclusion of mild AGN contamination in the regime of $-23 < M_{\rm UV} < -22$ inflating the LF and highlighting the need to properly handle AGN when measuring the LBG LF.
    \item Our best-fit values for both Schechter and DPL functional forms of the LBG LF are found to be consistent with the $z \sim 5$ measurements from \citet{Bowler2015} and the mild linear evolution models of the LBG LF from \citet{Bouwens2015}. These findings suggest that the shape of the bright-end of the UV LBG LF does not evolve significantly in the redshift range $3 < z < 5$.
    \item We suggest that discrepancies found between studies at the magnitude range where the LBG LF transitions into the AGN LF ($-24 < M_{\rm UV} < -23$) can be explained through differing definitions and methods of measuring $M_{\rm UV}$. Our proposed method of measuring $M_{\rm UV}$ through the use of the best fit SED to photometry and a thin top-hat filter positioned at the rest-frame 1500\AA \, allows for a more robust and consistent determination of $M_{\rm UV}$ for each object that is not reliant on a single measurement nor impacted by the combination of redshift and UV slope of the galaxy spectrum on broad-band filters.
    \item We measure the transition between AGN/LBG domination in the UV LF and find that regardless of the functional form used to fit for the LBGs, the 50 per cent AGN fraction occurs within the range of $-23.5 < M_{\rm UV} < -23$, in agreement with recent results from \citet{Boutsia2018}.
    \item We find agreement with recent studies suggesting a steep faint-end slope for the AGN UV LF at $z \sim 4$ with $\alpha_{AGN} = -2.09^{+0.35}_{-0.38}$ ($-1.66^{+0.29}_{-0.58}$) when fit simultaneously with a Schechter (DPL) for the LBGs. These results support the  conclusion from \citet{Stevans2018} whom suggest that AGN, while not the dominant source of ionising photons, could sustain re-ionisation of Hydrogen on their own at this epoch. Our results highlight the importance of simultaneously fitting the two populations of LBG and AGN together. Future insight into the nature of the sources at the transition (e.g. spectroscopic follow up with~\emph{VLT, JWST}) will shed light onto the astrophysics at play in shaping the bright end of the galaxy population.
\end{enumerate}

\section*{Acknowledgements}

The authors would like to pass on our thanks to J.Patterson and the University of Oxford's IT team at the Physics Department for their continued efforts. We thank the anonymous referee for useful comments that improved that paper. We also thank M.L.Stevans and S.L.Finkelstein for providing their final results for our comparisons.   We give additional thanks to the HSC team for compiling spectroscopic catalogues from a vast range of surveys and granting easy public access to those catalogues. This research  made  use  of Astropy,  a  community-developed core  Python  package  for  Astronomy  (Astropy  Collaboration,  2013).

NA acknowledges funding from the Science and Technology Facilities Council (STFC) Grant Code ST/R505006/1. This work was supported by the Glasstone Foundation,the Oxford Hintze Centre for Astrophysical Surveys which is funded through generous support from the Hintze Family Charitable Foundation and the award of the STFC consolidated grant (ST/N000919/1). 

This work is based on data products from observations made with ESO Telescopes at the La Silla Paranal Observatory under ESO programme ID 179.A-2005 and ID 179.A-2006 and on data products produced by CALET and the Cambridge Astronomy Survey Unit on behalf of the UltraVISTA and VIDEO consortia.

Based on observations obtained with MegaPrime/MegaCam, a joint project of CFHT and CEA/IRFU, at the Canada-France-Hawaii Telescope (CFHT) which is operated by the National Research Council (NRC) of Canada, the Institut National des Science de l'Univers of the Centre National de la Recherche Scientifique (CNRS) of France, and the University of Hawaii. This work is based in part on data products produced at Terapix available at the Canadian Astronomy Data Centre as part of the Canada-France-Hawaii Telescope Legacy Survey, a collaborative project of NRC and CNRS.

The Hyper Suprime-Cam (HSC) collaboration includes the astronomical communities of Japan and Taiwan, and Princeton University. The HSC instrumentation and software were developed by the National Astronomical Observatory of Japan (NAOJ), the Kavli Institute for the Physics and Mathematics of the Universe (Kavli IPMU), the University of Tokyo, the High Energy Accelerator Research Organization (KEK), the Academia Sinica Institute for Astronomy and Astrophysics in Taiwan (ASIAA), and Princeton University. Funding was contributed by the FIRST program from Japanese Cabinet Office, the Ministry of Education, Culture, Sports, Science and Technology (MEXT), the Japan Society for the Promotion of Science (JSPS), Japan Science and Technology Agency (JST), the Toray Science Foundation, NAOJ, Kavli IPMU, KEK, ASIAA, and Princeton University.

This paper makes use of software developed for the Large Synoptic Survey Telescope. We thank the LSST Project for making their code available as free software at  http://dm.lsst.org

This paper is based, in part, on data collected at the Subaru Telescope and retrieved from the HSC data archive system, which is operated by Subaru Telescope and Astronomy Data Center at National Astronomical Observatory of Japan. Data analysis was in part carried out with the cooperation of Center for Computational Astrophysics, National Astronomical Observatory of Japan.




\bibliographystyle{mnras}
\bibliography{mnras_template} 

\begin{thebibliography}{}
\makeatletter
\relax
\def\mn@urlcharsother{\let\do\@makeother \do\$\do\&\do\#\do\^\do\_\do\%\do\~}
\def\mn@doi{\begingroup\mn@urlcharsother \@ifnextchar [ {\mn@doi@}
  {\mn@doi@[]}}
\def\mn@doi@[#1]#2{\def\@tempa{#1}\ifx\@tempa\@empty \href
  {http://dx.doi.org/#2} {doi:#2}\else \href {http://dx.doi.org/#2} {#1}\fi
  \endgroup}
\def\mn@eprint#1#2{\mn@eprint@#1:#2::\@nil}
\def\mn@eprint@arXiv#1{\href {http://arxiv.org/abs/#1} {{\tt arXiv:#1}}}
\def\mn@eprint@dblp#1{\href {http://dblp.uni-trier.de/rec/bibtex/#1.xml}
  {dblp:#1}}
\def\mn@eprint@#1:#2:#3:#4\@nil{\def\@tempa {#1}\def\@tempb {#2}\def\@tempc
  {#3}\ifx \@tempc \@empty \let \@tempc \@tempb \let \@tempb \@tempa \fi \ifx
  \@tempb \@empty \def\@tempb {arXiv}\fi \@ifundefined
  {mn@eprint@\@tempb}{\@tempb:\@tempc}{\expandafter \expandafter \csname
  mn@eprint@\@tempb\endcsname \expandafter{\@tempc}}}

\bibitem[\protect\citeauthoryear{Adelberger \& Steidel}{Adelberger \&
  Steidel}{2000}]{Adelberger2000}
Adelberger K.~L.,  Steidel C.~C.,  2000, \mn@doi [\apj] {10.1086/317183}, 544,
  218

\bibitem[\protect\citeauthoryear{{Aihara} et~al.,}{{Aihara}
  et~al.}{2018a}]{Aihara2017}
{Aihara} H.,  et~al., 2018a, \mn@doi [\pasj] {10.1093/pasj/psx066}, \href
  {https://ui.adsabs.harvard.edu/abs/2018PASJ...70S...4A} {70, S4}

\bibitem[\protect\citeauthoryear{{Aihara} et~al.,}{{Aihara}
  et~al.}{2018b}]{Aihara2014}
{Aihara} H.,  et~al., 2018b, \mn@doi [\pasj] {10.1093/pasj/psx081}, \href
  {https://ui.adsabs.harvard.edu/abs/2018PASJ...70S...8A} {70, S8}

\bibitem[\protect\citeauthoryear{{Akiyama} et~al.,}{{Akiyama}
  et~al.}{2018}]{Akiyama2018}
{Akiyama} M.,  et~al., 2018, \mn@doi [\pasj] {10.1093/pasj/psx091}, \href
  {https://ui.adsabs.harvard.edu/abs/2018PASJ...70S..34A} {70, S34}

\bibitem[\protect\citeauthoryear{Alam et~al.,}{Alam et~al.}{2015}]{Alam2015}
Alam S.,  et~al., 2015, \mn@doi [\apjs] {10.1088/0067-0049/219/1/12}, 219, 12

\bibitem[\protect\citeauthoryear{{Arnouts}, {Cristiani}, {Moscardini},
  {Matarrese}, {Lucchin}, {Fontana}  \& {Giallongo}}{{Arnouts}
  et~al.}{1999}]{Arnouts1999}
{Arnouts} S.,  {Cristiani} S.,  {Moscardini} L.,  {Matarrese} S.,  {Lucchin}
  F.,  {Fontana} A.,   {Giallongo} E.,  1999, \mn@doi [\mnras]
  {10.1046/j.1365-8711.1999.02978.x}, \href
  {https://ui.adsabs.harvard.edu/abs/1999MNRAS.310..540A} {310, 540}

\bibitem[\protect\citeauthoryear{{Baldwin}, {Phillips}  \&
  {Terlevich}}{{Baldwin} et~al.}{1981}]{Baldwin1981}
{Baldwin} J.~A.,  {Phillips} M.~M.,   {Terlevich} R.,  1981, \mn@doi [\pasp]
  {10.1086/130766}, \href
  {https://ui.adsabs.harvard.edu/abs/1981PASP...93....5B} {93, 5}

\bibitem[\protect\citeauthoryear{{Begelman}}{{Begelman}}{2004}]{Begelman2003}
{Begelman} M.~C.,  2004, in {Ho} L.~C.,  ed., Coevolution of Black Holes and
  Galaxies. p.~374 (\mn@eprint {arXiv} {astro-ph/0303040})

\bibitem[\protect\citeauthoryear{{Benson}, {Bower}, {Frenk}, {Lacey}, {Baugh}
  \& {Cole}}{{Benson} et~al.}{2003}]{Benson2003}
{Benson} A.~J.,  {Bower} R.~G.,  {Frenk} C.~S.,  {Lacey} C.~G.,  {Baugh} C.~M.,
    {Cole} S.,  2003, \mn@doi [\apj] {10.1086/379160}, \href
  {https://ui.adsabs.harvard.edu/abs/2003ApJ...599...38B} {599, 38}

\bibitem[\protect\citeauthoryear{{Bertin}}{{Bertin}}{2011}]{Bertin2011}
{Bertin} E.,  2011, in {Evans} I.~N.,  {Accomazzi} A.,  {Mink} D.~J.,   {Rots}
  A.~H.,  eds,  ASP Conference Series Vol. 442, Astronomical Data Analysis
  Software and Systems XX. p.~435

\bibitem[\protect\citeauthoryear{Bertin \& Arnouts}{Bertin \&
  Arnouts}{1996}]{Bertin1996}
Bertin E.,  Arnouts S.,  1996, \aaps, 117, 393

\bibitem[\protect\citeauthoryear{Bian et~al.,}{Bian et~al.}{2013}]{Bian2013}
Bian F.,  et~al., 2013, \mn@doi [\apj] {10.1088/0004-637X/774/1/28}, 774, 28

\bibitem[\protect\citeauthoryear{Binney}{Binney}{2004}]{Binney2004}
Binney J.,  2004, \mn@doi [\mnras] {10.1111/j.1365-2966.2004.07277.x}, 347,
  1093

\bibitem[\protect\citeauthoryear{Bohlin, Colina  \& Finley}{Bohlin
  et~al.}{1995}]{Bohlin1995}
Bohlin R.,  Colina L.,   Finley D.,  1995, \apj, 110, 1316

\bibitem[\protect\citeauthoryear{Boutsia, Grazian, Giallongo, Fiore  \&
  Civano}{Boutsia et~al.}{2018}]{Boutsia2018}
Boutsia K.,  Grazian A.,  Giallongo E.,  Fiore F.,   Civano F.,  2018, \mn@doi
  [\apj] {10.3847/1538-4357/aae6c7}, 869, 20

\bibitem[\protect\citeauthoryear{Bouwens et~al.,}{Bouwens
  et~al.}{2015}]{Bouwens2015}
Bouwens R.~J.,  et~al., 2015, \mn@doi [\apj] {10.1088/0004-637X/803/1/34}, 803,
  1

\bibitem[\protect\citeauthoryear{Bower, Benson, Malbon, Helly, Frenk, Baugh,
  Cole  \& Lacey}{Bower et~al.}{2006}]{Bower2006}
Bower R.~G.,  Benson A.~J.,  Malbon R.,  Helly J.~C.,  Frenk C.~S.,  Baugh
  C.~M.,  Cole S.,   Lacey C.~G.,  2006, \mn@doi [\mnras]
  {10.1111/j.1365-2966.2006.10519.x}, 370, 645

\bibitem[\protect\citeauthoryear{Bowler et~al.,}{Bowler
  et~al.}{2014}]{Bowler2014}
Bowler R.~A.,  et~al., 2014, \mn@doi [\mnras] {10.1093/mnras/stu449}, 440, 2810

\bibitem[\protect\citeauthoryear{Bowler et~al.,}{Bowler
  et~al.}{2015}]{Bowler2015}
Bowler R.~A.,  et~al., 2015, \mn@doi [\mnras] {10.1093/mnras/stv1403}, 452,
  1817

\bibitem[\protect\citeauthoryear{{Bowler}, {Jarvis}, {Dunlop}, {McLure},
  {McLeod}, {Adams}, {Milvang-Jensen}  \& {McCracken}}{{Bowler}
  et~al.}{2019}]{bowler2019lack}
{Bowler} R.~A.~A.,  {Jarvis} M.~J.,  {Dunlop} J.~S.,  {McLure} R.~J.,  {McLeod}
  D.~J.,  {Adams} N.~J.,  {Milvang-Jensen} B.,   {McCracken} H.~J.,  2019,
  arXiv pre-print, \href
  {https://ui.adsabs.harvard.edu/abs/2019arXiv191112832B} {p. arXiv:1911.12832}

\bibitem[\protect\citeauthoryear{Bruzual \& Charlot}{Bruzual \&
  Charlot}{2003}]{Bruzual2003}
Bruzual G.,  Charlot S.,  2003, \mn@doi [\mnras]
  {10.1046/j.1365-8711.2003.06897.x}, 344, 1000

\bibitem[\protect\citeauthoryear{{Bunker} et~al.,}{{Bunker}
  et~al.}{2010}]{Bunker2010}
{Bunker} A.~J.,  et~al., 2010, \mn@doi [\mnras]
  {10.1111/j.1365-2966.2010.17350.x}, \href
  {https://ui.adsabs.harvard.edu/abs/2010MNRAS.409..855B} {409, 855}

\bibitem[\protect\citeauthoryear{{Calzetti}, {Armus}, {Bohlin}, {Kinney},
  {Koornneef}  \& {Storchi-Bergmann}}{{Calzetti} et~al.}{2000}]{Calzetti2000}
{Calzetti} D.,  {Armus} L.,  {Bohlin} R.~C.,  {Kinney} A.~L.,  {Koornneef} J.,
   {Storchi-Bergmann} T.,  2000, \mn@doi [\apj] {10.1086/308692}, \href
  {https://ui.adsabs.harvard.edu/abs/2000ApJ...533..682C} {533, 682}

\bibitem[\protect\citeauthoryear{{Chabrier}, {Baraffe}, {Allard}  \&
  {Hauschildt}}{{Chabrier} et~al.}{2000}]{Chabrier2000}
{Chabrier} G.,  {Baraffe} I.,  {Allard} F.,   {Hauschildt} P.,  2000, \mn@doi
  [\apj] {10.1086/309513}, \href
  {https://ui.adsabs.harvard.edu/abs/2000ApJ...542..464C} {542, 464}

\bibitem[\protect\citeauthoryear{{Ciotti} \& {Ostriker}}{{Ciotti} \&
  {Ostriker}}{1997}]{Ciotti1997}
{Ciotti} L.,  {Ostriker} J.~P.,  1997, \mn@doi [\apjl] {10.1086/310902}, \href
  {https://ui.adsabs.harvard.edu/abs/1997ApJ...487L.105C} {487, L105}

\bibitem[\protect\citeauthoryear{Coil et~al.,}{Coil et~al.}{2011}]{Coil2011}
Coil A.~L.,  et~al., 2011, \mn@doi [\apj] {10.1088/0004-637X/741/1/8}, 741, 8

\bibitem[\protect\citeauthoryear{Cole, Lacey, Baugh  \& Frenk}{Cole
  et~al.}{2002}]{Cole2002}
Cole S.,  Lacey C.~G.,  Baugh C.~M.,   Frenk C.~S.,  2002, \mn@doi [\mnras]
  {10.1046/j.1365-8711.2000.03879.x}, 319, 168

\bibitem[\protect\citeauthoryear{Conselice}{Conselice}{2014}]{Conselice2014}
Conselice C.~J.,  2014, \mn@doi [\araa] {10.1146/annurev-astro-081913-040037},
  52, 291

\bibitem[\protect\citeauthoryear{Cool et~al.,}{Cool et~al.}{2013}]{Cool2013}
Cool R.~J.,  et~al., 2013, \mn@doi [\apj] {10.1088/0004-637X/767/2/118}, 767,
  118

\bibitem[\protect\citeauthoryear{{Dekel} \& {Silk}}{{Dekel} \&
  {Silk}}{1986}]{Dekel1986}
{Dekel} A.,  {Silk} J.,  1986, \mn@doi [\apj] {10.1086/164050}, \href
  {https://ui.adsabs.harvard.edu/abs/1986ApJ...303...39D} {303, 39}

\bibitem[\protect\citeauthoryear{{Ferrarese} \& {Merritt}}{{Ferrarese} \&
  {Merritt}}{2000}]{Ferrarese2000}
{Ferrarese} L.,  {Merritt} D.,  2000, \mn@doi [\apjl] {10.1086/312838}, \href
  {https://ui.adsabs.harvard.edu/abs/2000ApJ...539L...9F} {539, L9}

\bibitem[\protect\citeauthoryear{Finkelstein et~al.,}{Finkelstein
  et~al.}{2015}]{Finkelstein2015}
Finkelstein S.~L.,  et~al., 2015, \mn@doi [\apj] {10.1088/0004-637X/810/1/71},
  810, 71

\bibitem[\protect\citeauthoryear{Giallongo et~al.,}{Giallongo
  et~al.}{2015}]{Giallongo2015}
Giallongo E.,  et~al., 2015, \mn@doi [\aap] {10.1051/0004-6361/201425334}, 578,
  A83

\bibitem[\protect\citeauthoryear{Glikman, Djorgovski, Stern, Dey, Jannuzi  \&
  Lee}{Glikman et~al.}{2011}]{Glikman2011}
Glikman E.,  Djorgovski S.~G.,  Stern D.,  Dey A.,  Jannuzi B.~T.,   Lee K.~S.,
   2011, \mn@doi [\apjl] {10.1088/2041-8205/728/2/L26}, 728, L26

\bibitem[\protect\citeauthoryear{Graham, Onken, Athanassoula  \& Combes}{Graham
  et~al.}{2011}]{Graham2011}
Graham A.~W.,  Onken C.~A.,  Athanassoula E.,   Combes F.,  2011, \mn@doi
  [\mnras] {10.1111/j.1365-2966.2010.18045.x}, 412, 2211

\bibitem[\protect\citeauthoryear{{Guhathakurta}, {Tyson}  \&
  {Majewski}}{{Guhathakurta} et~al.}{1990}]{Guhathakurta1990}
{Guhathakurta} P.,  {Tyson} J.~A.,   {Majewski} S.~R.,  1990, \mn@doi [\apjl]
  {10.1086/185754}, \href
  {https://ui.adsabs.harvard.edu/abs/1990ApJ...357L...9G} {357, L9}

\bibitem[\protect\citeauthoryear{Hamuy, Walker, Suntzeff, Gigoux, Heathcote  \&
  Phillips}{Hamuy et~al.}{1992}]{Hamuy1992}
Hamuy M.,  Walker A.,  Suntzeff N.,  Gigoux P.,  Heathcote S.,   Phillips M.,
  1992, \pasp, 104, 533

\bibitem[\protect\citeauthoryear{Hamuy, Sunteff, Heathcote, Walker, Gigoux  \&
  Phillips}{Hamuy et~al.}{1994}]{Hamuy1994}
Hamuy M.,  Sunteff N.,  Heathcote S.,  Walker A.,  Gigoux P.,   Phillips M.,
  1994, \pasp, 106, 566

\bibitem[\protect\citeauthoryear{{Hasinger} et~al.,}{{Hasinger}
  et~al.}{2018}]{Hasinger2018}
{Hasinger} G.,  et~al., 2018, \mn@doi [\apj] {10.3847/1538-4357/aabacf}, \href
  {https://ui.adsabs.harvard.edu/abs/2018ApJ...858...77H} {858, 77}

\bibitem[\protect\citeauthoryear{Hoaglin, Mosteller  \& Tukey}{Hoaglin
  et~al.}{1983}]{hoaglin2000understanding}
Hoaglin D.~C.,  Mosteller F.,   Tukey J.~W.,  1983, Understanding Robust and
  Exploratory Data Analysis (New York: Wiley)

\bibitem[\protect\citeauthoryear{Huang, Ferguson, Ravindranath  \& Su}{Huang
  et~al.}{2013}]{Huang2013}
Huang K.~H.,  Ferguson H.~C.,  Ravindranath S.,   Su J.,  2013, \mn@doi [\apj]
  {10.1088/0004-637X/765/1/68}, 765, 68

\bibitem[\protect\citeauthoryear{{Ilbert} et~al.,}{{Ilbert}
  et~al.}{2006}]{Ilbert2006}
{Ilbert} O.,  et~al., 2006, \mn@doi [\aap] {10.1051/0004-6361:20065138}, \href
  {https://ui.adsabs.harvard.edu/abs/2006A&A...457..841I} {457, 841}

\bibitem[\protect\citeauthoryear{Ilbert et~al.,}{Ilbert
  et~al.}{2009}]{Ilbert2009}
Ilbert O.,  et~al., 2009, \mn@doi [\apj] {10.1088/0004-637X/690/2/1236}, 690,
  1236

\bibitem[\protect\citeauthoryear{{Inoue}, {Shimizu}, {Iwata}  \&
  {Tanaka}}{{Inoue} et~al.}{2014}]{Inoue2014}
{Inoue} A.~K.,  {Shimizu} I.,  {Iwata} I.,   {Tanaka} M.,  2014, \mn@doi
  [\mnras] {10.1093/mnras/stu936}, \href
  {https://ui.adsabs.harvard.edu/abs/2014MNRAS.442.1805I} {442, 1805}

\bibitem[\protect\citeauthoryear{Jarvis et~al.,}{Jarvis
  et~al.}{2013}]{Jarvis2013}
Jarvis M.~J.,  et~al., 2013, \mn@doi [\mnras] {10.1093/mnras/sts118}, 428, 1281

\bibitem[\protect\citeauthoryear{{Kauffmann}, {Colberg}, {Diaferio}  \&
  {White}}{{Kauffmann} et~al.}{1999}]{Kauffmann1999}
{Kauffmann} G.,  {Colberg} J.~M.,  {Diaferio} A.,   {White} S. D.~M.,  1999,
  \mn@doi [\mnras] {10.1046/j.1365-8711.1999.02202.x}, \href
  {https://ui.adsabs.harvard.edu/abs/1999MNRAS.303..188K} {303, 188}

\bibitem[\protect\citeauthoryear{{Koratkar} \& {Blaes}}{{Koratkar} \&
  {Blaes}}{1999}]{Koratkar1999}
{Koratkar} A.,  {Blaes} O.,  1999, \mn@doi [\pasp] {10.1086/316294}, \href
  {https://ui.adsabs.harvard.edu/abs/1999PASP..111....1K} {111, 1}

\bibitem[\protect\citeauthoryear{Lawrence et~al.,}{Lawrence
  et~al.}{2007}]{Lawrence2007}
Lawrence A.,  et~al., 2007, \mn@doi [\mnras]
  {10.1111/j.1365-2966.2007.12040.x}, 379, 1599

\bibitem[\protect\citeauthoryear{LeF\`{e}vre et~al.,}{LeF\`{e}vre
  et~al.}{2013}]{LeFevre2013}
LeF\`{e}vre O.,  et~al., 2013, \mn@doi [\aap] {10.1051/0004-6361/201322179},
  559, A14

\bibitem[\protect\citeauthoryear{{Levenberg}}{{Levenberg}}{1944}]{Levenberg1944}
{Levenberg} K.,  1944, \mn@doi [Quart. Appl. Math.] {10.1090/10666}, pp
  164--168

\bibitem[\protect\citeauthoryear{Lilly et~al.,}{Lilly et~al.}{2009}]{Lilly2009}
Lilly S.~J.,  et~al., 2009, \mn@doi [\apjs] {10.1088/0067-0049/184/2/218}, 184,
  218

\bibitem[\protect\citeauthoryear{{Madau}}{{Madau}}{1995}]{Madau1995}
{Madau} P.,  1995, \mn@doi [\apj] {10.1086/175332}, \href
  {https://ui.adsabs.harvard.edu/abs/1995ApJ...441...18M} {441, 18}

\bibitem[\protect\citeauthoryear{{Marchesi} et~al.,}{{Marchesi}
  et~al.}{2016}]{Marchesi2016}
{Marchesi} S.,  et~al., 2016, \mn@doi [\apj] {10.3847/0004-637X/817/1/34},
  \href {https://ui.adsabs.harvard.edu/abs/2016ApJ...817...34M} {817, 34}

\bibitem[\protect\citeauthoryear{Marquardt}{Marquardt}{1963}]{marquardt:1963}
Marquardt D.~W.,  1963, \mn@doi [SIAM Journal on Applied Mathematics]
  {10.1137/0111030}, 11, 431

\bibitem[\protect\citeauthoryear{Masters et~al.,}{Masters
  et~al.}{2012}]{Masters2012}
Masters D.,  et~al., 2012, \mn@doi [\apj] {10.1088/0004-637X/755/2/169}, 755,
  169

\bibitem[\protect\citeauthoryear{McCracken et~al.,}{McCracken
  et~al.}{2012}]{McCracken2012}
McCracken H.~J.,  et~al., 2012, \mn@doi [\aap] {10.1051/0004-6361/201219507},
  544, A156

\bibitem[\protect\citeauthoryear{McLure, Cirasuolo, Dunlop, Foucaud  \&
  Almaini}{McLure et~al.}{2009}]{Mclure2009}
McLure R.~J.,  Cirasuolo M.,  Dunlop J.~S.,  Foucaud S.,   Almaini O.,  2009,
  \mn@doi [\mnras] {10.1111/j.1365-2966.2009.14677.x}, 395, 2196

\bibitem[\protect\citeauthoryear{McLure et~al.,}{McLure
  et~al.}{2013}]{McLure2013}
McLure R.~J.,  et~al., 2013, \mn@doi [\mnras] {10.1093/mnras/stt627}, 432, 2696

\bibitem[\protect\citeauthoryear{{McLure} et~al.,}{{McLure}
  et~al.}{2018}]{McLure2018}
{McLure} R.~J.,  et~al., 2018, \mn@doi [\mnras] {10.1093/mnras/sty1213}, \href
  {https://ui.adsabs.harvard.edu/abs/2018MNRAS.479...25M} {479, 25}

\bibitem[\protect\citeauthoryear{Momcheva et~al.,}{Momcheva
  et~al.}{2016}]{Momcheva2016}
Momcheva I.~G.,  et~al., 2016, \mn@doi [\apjs] {10.3847/0067-0049/225/2/27},
  225, 27

\bibitem[\protect\citeauthoryear{Nagamine, Fukugita, Cen  \& Ostriker}{Nagamine
  et~al.}{2001}]{Nagamine2001}
Nagamine K.,  Fukugita M.,  Cen R.,   Ostriker J.~P.,  2001, \mn@doi [\mnras]
  {10.1046/j.1365-8711.2001.04905.x}, 327, L10

\bibitem[\protect\citeauthoryear{Oesch et~al.,}{Oesch et~al.}{2012}]{Oesch2012}
Oesch P.~A.,  et~al., 2012, \mn@doi [\apj] {10.1088/0004-637X/759/2/135}, 759

\bibitem[\protect\citeauthoryear{{Oke}}{{Oke}}{1974}]{Oke1974}
{Oke} J.~B.,  1974, \mn@doi [\apjs] {10.1086/190287}, \href
  {https://ui.adsabs.harvard.edu/abs/1974ApJS...27...21O} {27, 21}

\bibitem[\protect\citeauthoryear{{Oke} \& {Gunn}}{{Oke} \&
  {Gunn}}{1983}]{Oke1983}
{Oke} J.~B.,  {Gunn} J.~E.,  1983, \mn@doi [\apj] {10.1086/160817}, \href
  {https://ui.adsabs.harvard.edu/abs/1983ApJ...266..713O} {266, 713}

\bibitem[\protect\citeauthoryear{{Ono} et~al.,}{{Ono} et~al.}{2018}]{Ono2017}
{Ono} Y.,  et~al., 2018, \mn@doi [\pasj] {10.1093/pasj/psx103}, \href
  {https://ui.adsabs.harvard.edu/abs/2018PASJ...70S..10O} {70, S10}

\bibitem[\protect\citeauthoryear{Papovich et~al.,}{Papovich
  et~al.}{2016}]{Papovich2016}
Papovich C.,  et~al., 2016, \mn@doi [\apjs] {10.3847/0067-0049/224/2/28}, 224,
  28

\bibitem[\protect\citeauthoryear{{Parsa}, {Dunlop}  \& {McLure}}{{Parsa}
  et~al.}{2018}]{Parsa2018}
{Parsa} S.,  {Dunlop} J.~S.,   {McLure} R.~J.,  2018, \mn@doi [\mnras]
  {10.1093/mnras/stx2887}, \href
  {https://ui.adsabs.harvard.edu/abs/2018MNRAS.474.2904P} {474, 2904}

\bibitem[\protect\citeauthoryear{{Pentericci} et~al.,}{{Pentericci}
  et~al.}{2018}]{Pentericci2018}
{Pentericci} L.,  et~al., 2018, \mn@doi [\aap] {10.1051/0004-6361/201833047},
  \href {https://ui.adsabs.harvard.edu/abs/2018A&A...616A.174P} {616, A174}

\bibitem[\protect\citeauthoryear{{Pickles}}{{Pickles}}{1998}]{Pickles1998}
{Pickles} A.~J.,  1998, \mn@doi [\pasp] {10.1086/316197}, \href
  {https://ui.adsabs.harvard.edu/abs/1998PASP..110..863P} {110, 863}

\bibitem[\protect\citeauthoryear{{Polletta} et~al.,}{{Polletta}
  et~al.}{2007}]{Polletta2007}
{Polletta} M.,  et~al., 2007, \mn@doi [\apj] {10.1086/518113}, \href
  {https://ui.adsabs.harvard.edu/abs/2007ApJ...663...81P} {663, 81}

\bibitem[\protect\citeauthoryear{Powell, Slyz  \& Devriendt}{Powell
  et~al.}{2011}]{Powell2011}
Powell L.~C.,  Slyz A.,   Devriendt J.,  2011, \mn@doi [\mnras]
  {10.1111/j.1365-2966.2011.18668.x}, 414, 3671

\bibitem[\protect\citeauthoryear{Ricci, Marchesi, Shankar, Franca  \&
  Civano}{Ricci et~al.}{2017}]{Ricci2017}
Ricci F.,  Marchesi S.,  Shankar F.,  Franca F.~L.,   Civano F.,  2017, \mn@doi
  [\mnras] {10.1093/mnras/stw2909}, 465, 1915

\bibitem[\protect\citeauthoryear{{Rowan-Robinson}}{{Rowan-Robinson}}{1968}]{rowanrobinson1968}
{Rowan-Robinson} M.,  1968, \mn@doi [\mnras] {10.1093/mnras/138.4.445}, \href
  {https://ui.adsabs.harvard.edu/abs/1968MNRAS.138..445R} {138, 445}

\bibitem[\protect\citeauthoryear{Salvato et~al.,}{Salvato
  et~al.}{2009}]{Salvato2009}
Salvato M.,  et~al., 2009, \mn@doi [\apj] {10.1088/0004-637X/690/2/1250}, 690,
  1250

\bibitem[\protect\citeauthoryear{Schawinski, Thomas, Sarzi, Maraston, Kaviraj,
  Joo, Yi  \& Silk}{Schawinski et~al.}{2007}]{Schawinski2007}
Schawinski K.,  Thomas D.,  Sarzi M.,  Maraston C.,  Kaviraj S.,  Joo S.~J.,
  Yi S.~K.,   Silk J.,  2007, \mn@doi [\mnras]
  {10.1111/j.1365-2966.2007.12487.x}, 382, 1415

\bibitem[\protect\citeauthoryear{{Schechter}}{{Schechter}}{1976}]{Schechter1976}
{Schechter} P.,  1976, \mn@doi [\apj] {10.1086/154079}, \href
  {https://ui.adsabs.harvard.edu/abs/1976ApJ...203..297S} {203, 297}

\bibitem[\protect\citeauthoryear{Schenker et~al.,}{Schenker
  et~al.}{2013}]{Schenker2013}
Schenker M.~A.,  et~al., 2013, \mn@doi [\apj] {10.1088/0004-637X/768/2/196},
  768

\bibitem[\protect\citeauthoryear{{Schmidt}}{{Schmidt}}{1968}]{Schmidt1968}
{Schmidt} M.,  1968, \mn@doi [\apj] {10.1086/149446}, \href
  {https://ui.adsabs.harvard.edu/abs/1968ApJ...151..393S} {151, 393}

\bibitem[\protect\citeauthoryear{Schmidt et~al.,}{Schmidt
  et~al.}{2014}]{Schmidt2014}
Schmidt K.~B.,  et~al., 2014, \mn@doi [\apj] {10.1088/0004-637X/786/1/57}, 786

\bibitem[\protect\citeauthoryear{{S{\'e}rsic}}{{S{\'e}rsic}}{1963}]{1963BAAA....6...41S}
{S{\'e}rsic} J.~L.,  1963, Boletin de la Asociacion Argentina de Astronomia La
  Plata Argentina, \href
  {https://ui.adsabs.harvard.edu/abs/1963BAAA....6...41S} {6, 41}

\bibitem[\protect\citeauthoryear{Silk \& Mamon}{Silk \& Mamon}{2012}]{Silk2012}
Silk J.,  Mamon G.~A.,  2012, \mn@doi [Research in Astronomy and Astrophysics]
  {10.1088/1674-4527/12/8/004}, 12, 917

\bibitem[\protect\citeauthoryear{{Silk} \& {Rees}}{{Silk} \&
  {Rees}}{1998}]{Silk1998}
{Silk} J.,  {Rees} M.~J.,  1998, \aap, \href
  {https://ui.adsabs.harvard.edu/abs/1998A&A...331L...1S} {331, L1}

\bibitem[\protect\citeauthoryear{{Silva}, {Granato}, {Bressan}  \&
  {Danese}}{{Silva} et~al.}{1998}]{Silva1998}
{Silva} L.,  {Granato} G.~L.,  {Bressan} A.,   {Danese} L.,  1998, \mn@doi
  [\apj] {10.1086/306476}, \href
  {https://ui.adsabs.harvard.edu/abs/1998ApJ...509..103S} {509, 103}

\bibitem[\protect\citeauthoryear{Silverman et~al.,}{Silverman
  et~al.}{2015}]{Silverman2015}
Silverman J.~D.,  et~al., 2015, \mn@doi [\apjs] {10.1088/0067-0049/220/1/12},
  220, 12

\bibitem[\protect\citeauthoryear{Skelton et~al.,}{Skelton
  et~al.}{2014}]{Skelton2014}
Skelton R.~E.,  et~al., 2014, \mn@doi [\apjs] {10.1088/0067-0049/214/2/24},
  214, 24

\bibitem[\protect\citeauthoryear{{Steidel} \& {Hamilton}}{{Steidel} \&
  {Hamilton}}{1992}]{Steidel1992}
{Steidel} C.~C.,  {Hamilton} D.,  1992, \mn@doi [\aj] {10.1086/116287}, \href
  {https://ui.adsabs.harvard.edu/abs/1992AJ....104..941S} {104, 941}

\bibitem[\protect\citeauthoryear{{Steidel}, {Giavalisco}, {Pettini},
  {Dickinson}  \& {Adelberger}}{{Steidel} et~al.}{1996}]{Steidel1996}
{Steidel} C.~C.,  {Giavalisco} M.,  {Pettini} M.,  {Dickinson} M.,
  {Adelberger} K.~L.,  1996, \mn@doi [\apjl] {10.1086/310029}, \href
  {https://ui.adsabs.harvard.edu/abs/1996ApJ...462L..17S} {462, L17}

\bibitem[\protect\citeauthoryear{{Stevans} et~al.,}{{Stevans}
  et~al.}{2018}]{Stevans2018}
{Stevans} M.~L.,  et~al., 2018, \mn@doi [\apj] {10.3847/1538-4357/aacbd7},
  \href {https://ui.adsabs.harvard.edu/abs/2018ApJ...863...63S} {863, 63}

\bibitem[\protect\citeauthoryear{{Trenti} \& {Stiavelli}}{{Trenti} \&
  {Stiavelli}}{2008}]{Trenti2008}
{Trenti} M.,  {Stiavelli} M.,  2008, \mn@doi [\apj] {10.1086/528674}, \href
  {https://ui.adsabs.harvard.edu/abs/2008ApJ...676..767T} {676, 767}

\bibitem[\protect\citeauthoryear{{Viironen} et~al.,}{{Viironen}
  et~al.}{2018}]{Viironen2018}
{Viironen} K.,  et~al., 2018, \mn@doi [\aap] {10.1051/0004-6361/201731797},
  \href {https://ui.adsabs.harvard.edu/abs/2018A&A...614A.129V} {614, A129}

\bibitem[\protect\citeauthoryear{Wang \& Heckman}{Wang \&
  Heckman}{1996}]{Wang1996}
Wang B.,  Heckman T.,  1996, \apj, 457, 645

\bibitem[\protect\citeauthoryear{{Wilkins}, {Bunker}, {Lorenzoni}  \&
  {Caruana}}{{Wilkins} et~al.}{2011}]{Wilkins2011}
{Wilkins} S.~M.,  {Bunker} A.~J.,  {Lorenzoni} S.,   {Caruana} J.,  2011,
  \mn@doi [\mnras] {10.1111/j.1365-2966.2010.17626.x}, \href
  {https://ui.adsabs.harvard.edu/abs/2011MNRAS.411...23W} {411, 23}

\bibitem[\protect\citeauthoryear{van~der Burg, Hildebrandt  \& Erben}{van~der
  Burg et~al.}{2010}]{vdb2010}
van~der Burg R. F.~J.,  Hildebrandt H.,   Erben T.,  2010, \mn@doi [\aap]
  {10.1051/0004-6361/200913812}, 523, A74

\makeatother
\end{thebibliography}


\appendix

\section{Discrepancies in the Transition Between AGN and LBG Dominance}
We show here the impact of measuring the transition between AGN and LBG dominance with two differing methods of defining $M_{\textrm {UV}}$. One with a top-hat function centered on 1500\AA\, of the rest-frame best-fit SED for each object and the other method simply using the averaged $i$-band flux from each object. Here we find that when using the $i$-band flux to directly translate to an absolute UV magnitude that our results closely match those of \citet{Stevans2018}, while when using the top-hat function the number densities decrease.

\begin{figure}
    \centering
    \includegraphics[width=\columnwidth]{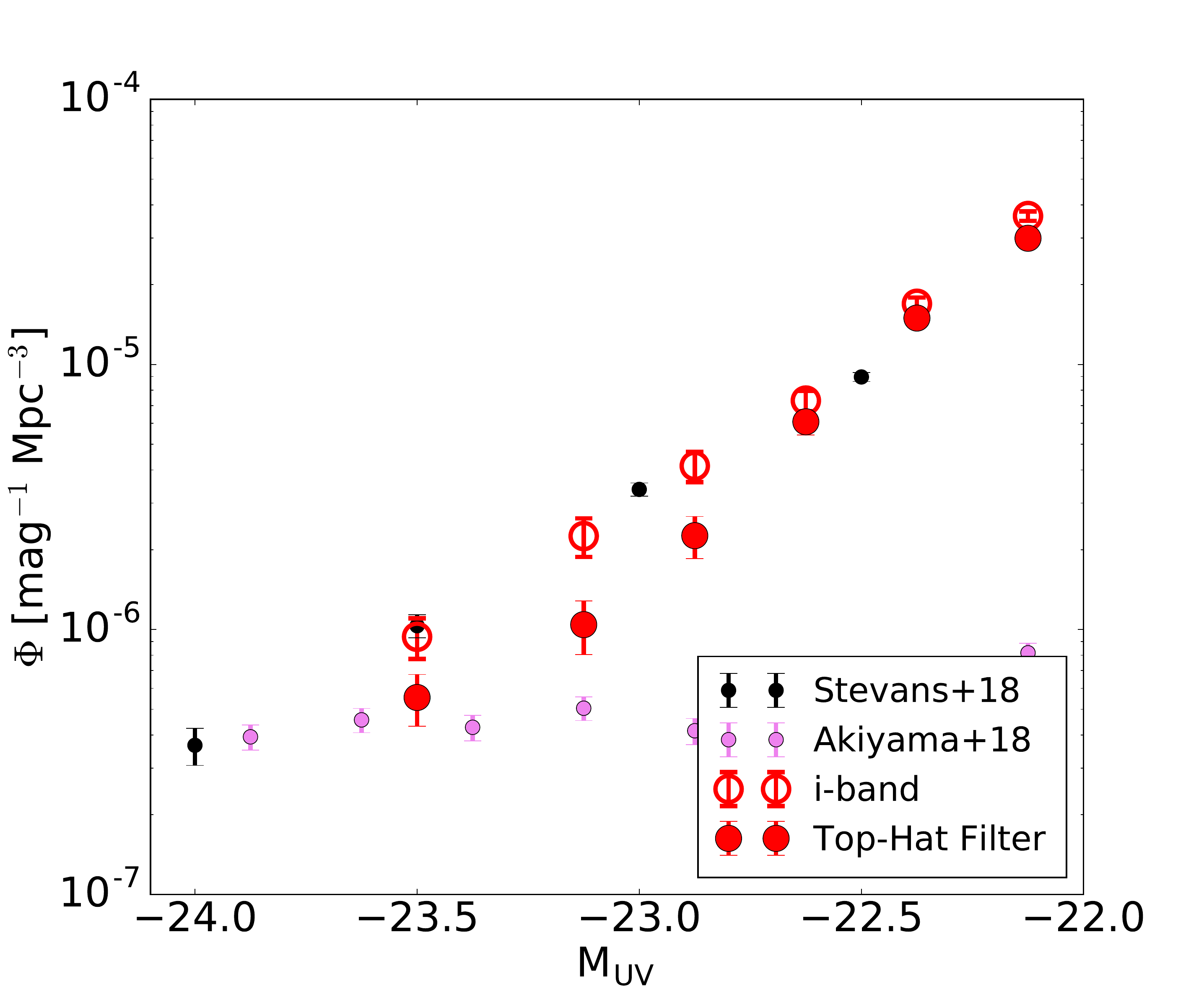}
    \caption{A zoom into the transition region at $-24 < M_{\rm UV} < -23$ showing how the UV LF changes depending on the method of measuring $M_{\textrm{UV}}$. We note that when using the average $i$-band measurement that our results closely match those of \citet{Stevans2018}. Also shown are the data from \citet{Akiyama2018} to show how the two methods converge onto the AGN LF in different places.}
    \label{fig:MUV}
\end{figure}

\vfill\eject
\section{Tabular data set}

Presented in Tab.~\ref{Tab:Points} are the binned rest-frame UV LF data points at $3.5 < z < 4.5$ as measured from this study.

\begin{table}
\caption{The rest-frame UV LF and its error margin at $3.5 < z < 4.5$. Column 1 shows the absolute UV magnitude at 1500\AA\, ($M_{\rm UV}$). Column 2 shows the number density of objects and column 3 shows the errors in the number density which are calculated with equation 2. Both the value of the number density and its corresponding error are in a base 10 logarithmic scale.}
\begin{tabular}{ccc}
\hline
$M_{\rm UV}$     & $\textrm{log}_{10}(\Phi)$ & $\delta \textrm{log}_{10}(\Phi)$ \\
$[\textrm{mag}]$ & $[\textrm{mag}^{-1} \textrm{Mpc}^{-3}]$ & $[\textrm{mag}^{-1} \textrm{Mpc}^{-3}]$ \\ \hline
$-27.250$ & $-7.792$ & 0.997 \\
$-26.250$ & $-7.330$ & 0.336 \\
$-25.250$ & $-6.949$ & 0.183 \\
$-24.250$ & $-6.511$ & 0.103 \\
$-23.500$ & $-6.256$ & 0.109 \\
$-23.125$ & $5.982 $& 0.112 \\
$-22.875$ & $5.646 $& 0.087 \\
$-22.625$ & $-5.216$ & 0.050 \\
$-22.375$ & $-4.825$ & 0.036 \\
$-22.125$ & $-4.524$ & 0.030 \\
$-21.875$ & $-4.160$ & 0.026 \\
$-21.625$ & $-3.861$ & 0.024 \\
$-21.375$ & $-3.588$ & 0.023 \\
$-21.125$ & $-3.391$ & 0.023 \\
$-20.875$ & $-3.219$ & 0.023 \\
$-20.625$ & $-3.095$ & 0.023 \\
$-20.375$ & $-2.975$ & 0.022 \\
$-20.125$ & $-2.848$ & 0.022 \\ \hline
\label{Tab:Points}
\end{tabular}
\end{table}



\bsp	
\label{lastpage}
\end{document}